\documentclass[ reprint,
nofootinbib,
 amsmath,amssymb,
 aps,
]{revtex4-2}

\usepackage[dvipsnames, usenames]{xcolor}
\usepackage{graphicx}
\usepackage[colorlinks]{hyperref}
\usepackage[nameinlink]{cleveref} 
\usepackage{bm}
\usepackage{comment}

\Crefname{equation}{}{}
\usepackage{xcolor}

\hypersetup{colorlinks=true,citecolor=black,filecolor=black,linkcolor=black,urlcolor=black}

\begin{document}

\title{Learning orbital dynamics of binary black hole systems from\\ gravitational wave measurements}
\author{Brendan Keith}\email[Corresponding author: ]{keith10@llnl.gov}
\affiliation{
Center for Applied Scientific Computing, Lawrence Livermore National Laboratory}
\affiliation{Institute for Computational and Experimental Research in Mathematics (ICERM), Brown University}
\author{Akshay Khadse}\affiliation{Department of Physics and Astronomy, The University of Mississippi, University, MS 38677, USA}
\author{Scott E. Field}\affiliation{Department of Mathematics, Center for Scientific Computing and Visualization Research, University of Massachusetts, Dartmouth, MA 02747, USA}

\begin{abstract}
We introduce a gravitational waveform inversion strategy that discovers mechanical models of binary black hole (BBH) systems. We show that only a single time series of (possibly noisy) waveform data is necessary to construct the equations of motion for a BBH system. Starting with a class of universal differential equations parameterized by feed-forward neural networks, our strategy involves the construction of a space of plausible mechanical models and a physics-informed constrained optimization within that space to minimize the waveform error.  We apply our method to various BBH systems including extreme and comparable mass ratio systems in eccentric and non-eccentric orbits. We show that the resulting differential equations apply to time durations longer than the training interval, and relativistic effects, such as perihelion precession, radiation reaction, and orbital plunge, are automatically accounted for. The methods outlined here provide a new, data-driven approach to studying the dynamics of binary black hole systems.
\end{abstract}
\maketitle

\section{Introduction} \label{sec:introduction}

Classical physical theories begin with scientific laws as ans\"atze, which are validated by repeated scientific experiment.
From these laws, one derives a set of equations (usually differential equations) that can be solved, either completely or partially, to deduce various conclusions about the physical system under consideration.
In this paper, we follow a different approach to 
learning physical equations: we solve an optimization problem that isolates the most likely physical model (differential equations) that would deliver certain physical measurements (data).
This approach is aligned with a growing trend in data-driven science; see, e.g., \cite{crutchfield1987equations,brunton2016discovering,raissi2018hidden,chen2018neural,long2018pde,han2019uniformly,wu2019numerical,rackauckas2020universal,huang2020learning,pmlr-v107-sun20a,yazdani2020systems,dandekar2020bayesian}.

The focus of this paper is on learning mechanical models for binary black hole (BBH) systems through gravitational wave measurements. 
As the  black holes orbit one another, the motion of these massive objects generate 
gravitational waves that radiate away to the far-field where they 
can be observed by an international-network of detectors~\cite{abbott2020gwtc}.
Complicated partial differential equations (PDEs) govern the entire process, and in particular
connect the near-field dynamics to the far-field gravitational radiation. Traditionally, 
black hole orbital dynamics and 
gravitational waves have been computed by expensive simulation codes~\cite{lehner2014numerical}
or approximations to general relativity such as the post-Newtonian formalism~\cite{blanchet2014gravitational}.

Our principal contribution is to show that two-body relativistic orbital models can be deduced from gravitational wave (GW) measurements 
by solving an inverse problem \cite{vogel2002computational,arridge_maass_oktem_schonlieb_2019} where the control variable is the vector of weights and biases in a neural network. Our numerical examples use
gravitational waveform measurements from both a noisy detector and
``clean" measurements from numerical relativity (NR)
simulations. As such, the techniques described here may apply to 
both traditional modeling endeavors that require NR data 
for calibration and GW astronomy. In the latter case,
where the training data is comprised of GW observations,
our inversion strategy avoids the need to solve Einstein's equation of general relativity
to learn the orbital model.

A key goal of our paper is to 
develop the computational framework for learning binary
black hole dynamical models from gravitational waves, 
which is a new approach to the modeling problem. As such, 
we focus on simple modeling choices and apply them
to illustrative examples. We will show that
simple ansatz models parameterized by feed-forward neural networks
can be used to discover complicated dynamics. Indeed, despite starting
with an essentially Newtonian ansatz model (cf. Eq.~\eqref{eq:UDEModel}),
our trained models accurately capture both the relativistic dynamics
and the waveform (cf. Sec.~\ref{sec:examples}). Potential applications are considered
in the Discussion section.

\section{Methodology} 
\subsection{Universal differential equations} \label{sec:universal_differential_equations}

In this work, we rely on the following general class of dynamical system models referred to in \cite{rackauckas2020universal} as (autonomous) universal differential equations (UDEs),
\begin{align}
\label{eq:UDE}
	\dot{\mathbf{x}}
	=
			\mathbf{f}(\mathbf{x},\mathcal{F}(\mathbf{x}))
	,
	\qquad
	\mathbf{x}(0) &= \mathbf{x}_0
		\,,
\end{align}
where $\mathbf{x}(t)$ is the solution vector and $\mathbf{x}_0$ specifies the initial conditions.
Here, $\mathcal{F}(\mathbf{x})$ is neural network and the overdot symbol ``$~\dot{}~$'' denotes differentiation with respect to time $t \in [0,T]$.
For example, if $\mathcal{F}(\mathbf{x}) = \mathcal{F}(\mathbf{x};\bm{\xi})$ is a feed-forward deep neural network with two hidden layers, it can be written as
\begin{equation}
\label{eq:TwoHiddenLayers}
	\mathcal{F}(\mathbf{x};\bm{\xi})
	=
	W_3\,\sigma_2(W_2\,\sigma_1(W_1(\mathbf{x}) + b_1)+b_2)+b_3 \,,
\end{equation}
with $\bm{\xi} = (W_1,W_2,W_3,b_1,b_2,b_3)$.
Here, $W_i$ are matrices (weights), $b_i$ are vectors (biases), and $\sigma_i$ are the chosen activation functions.

We note that the UDE paradigm permits a immense variety of different parameterized functions $\mathcal{F}(\mathbf{x})$, not only neural network-based parameterizations.
This flexibility provides numerous advantages, for example, prior scientific knowledge of the solution may be incorporated into the choice of function parametrization.
We choose to use feed-forward neural networks, such as~\cref{eq:TwoHiddenLayers}, because of the ease with which their parametrization can be determined using existing software \cite{rackauckas2020universal}.

\subsection{BBH modeling} \label{sec:physically_motivated_models}

Our task is to define a family of physical models  that can be used to describe
relativistic orbital dynamics of two spherical objects of mass $m_1$ and $m_2$. That is, we 
ask that our model provides the position of object 1, $\mathbf{r}_1(t)$, and 
object 2, $\mathbf{r}_2(t)$, which 
are the solutions to the dynamical system model. 
In Newtonian physics,
this is the familiar two-body problem of Kepler whose solution has been known since 
the earliest days of classical mechanics.
In relativistic physics, the field of 
computational relativity is largely devoted to providing numerical solutions to this
problem by solving the equations of general
relativity (a nonlinear, coupled, hyperbolic-elliptic PDE system) on
large supercomputers~\cite{lehner2014numerical}. In order to motivate 
a dynamical system model, we consider special cases whereby the 
dynamical motion described by PDEs can be 
approximately reduced to ordinary differential equations (ODEs). 
Such approximations have been well developed over many decades,
and we refer the reader
to~\cite{poisson2014gravity,blanchet2014gravitational,chandrasekhar1998mathematical} 
and references therein. Throughout this paper, 
we use geometric units where both the 
speed of light, $c$, and the gravitational constant, $G$,
are set to unity. 

First, for slow moving
objects ($v \ll c$), a powerful formalism
known as the post-Newtonian approximation
provides a systematic framework for 
adding relativistic corrections in powers of $v/c$ to the Keplerian equations of motion.
The post-Newtonian framework provides us with justification for treating the two-body problem
as an effective one-body problem. Here, the relevant equations that govern the separation
vector, $\mathbf{r} = \mathbf{r}_1 - \mathbf{r}_2$, can be used to reconstruct the two-body motion through
the relations,
\begin{align}
\label{eq:body_map}
\mathbf{r}_1 = (m_2/M) \mathbf{r} \,, \qquad \mathbf{r}_1 = -(m_1/M) \mathbf{r} \,,
\end{align}
where $M = m_1 + m_2$.
We call this an effective one-body problem as one often views there to be an ``effective" body
located at $\mathbf{r}$ relative to the system's center-of-mass.
A different technique, known as the post-Minkowskian approximation, informs us that the dominant 
contribution to the 
gravitational radiation field can be computed according to the quadrupole formula, first derived by Einstein. 
Collectively, the two-body-to-one-body map and the quadrupole formula neglect a
substantial amount of physics including terms proportional to 
$v / c$, as well as many more higher-order terms, some of which have been computed and others that
remain unknown.

Second, in the limit of $m_1 \gg m_2$, the two-body problem reduces to a
simpler setup whereby the larger
object is fixed at the coordinate system's origin. The smaller orbiting object's motion is then 
described by a geodesic path in the Schwarzschild geometry set by the 
larger black hole. Recent numerical evidence suggests
that the geodesic equations of motion~\eqref{eq:EMRIModel} 
with self-force corrections may work unreasonably well
even for near-equal mass systems~\cite{lewis2017fundamental,zimmerman2016redshift,le2011periastron,le2012gravitational,le2014overlap,le2013periastron,rifat2020surrogate,pound2019secondorder,van2020intermediate}. 
More importantly for our purposes, we know from 
blackhole perturbation theory results that using geodesic equations of motion 
to describe the two-body problem neglects a
substantial amount of physics including terms proportional to $m_2 / m_1$.

Having motivated some of the physics behind the problem, we now outline 
our strategy to write down a model
inspired by a combination of Newtonian and relativistic physics.
Our orbital model will omit a significant amount of important physics 
that will be accounted for by deep neural networks trained on 
gravitational waveform data.

We write the two-body problem
as an effective one-body one, and associate the orbital separation vector, $\mathbf{r}$,
with the location of the fictitious effective body orbiting a fixed, spherically-symmetric central object.
Owing to the spherical symmetry of the central object,
we may assume, without loss of generality, that the effective object's trajectory
lies in the equatorial plane, which we take to be 
the plane perpendicular to the z-axis and where the 
\emph{angle} $\phi$ is between $\mathbf{r}$ and the x-axis.
Orbits are specified by 
the orbital parameters 
{\em eccentricity} $e(t)$ and {\em semi--latus rectum} $p(t)$; in the Newtonian
case these are constants while in the general relativistic case they are 
often interpreted as time-dependent 
functions. Finally, we use a well-known parameterization for the Euclidean norm of $\mathbf{r}$,
\begin{align} \label{eq:r_parameterization}
r(t) = p(t) M / (1+e(t) \cos\chi(t)) \,,
\end{align}
and evolve the \emph{anomaly} $\chi(t)$
instead of $r(t)$ because the anomaly increases 
monotonically through radial turning points. To summarize, we assume the equations of motion
for the effective object can be described by four time-dependent variables
$\phi$, $\chi$, $p$, and $e$. The effective object's trajectory 
is provided $\phi,~\chi$ whereas $p,~e$ parameterize
the orbital configuration. 

\begin{figure*}
\centering
  \includegraphics[width=0.9\linewidth]{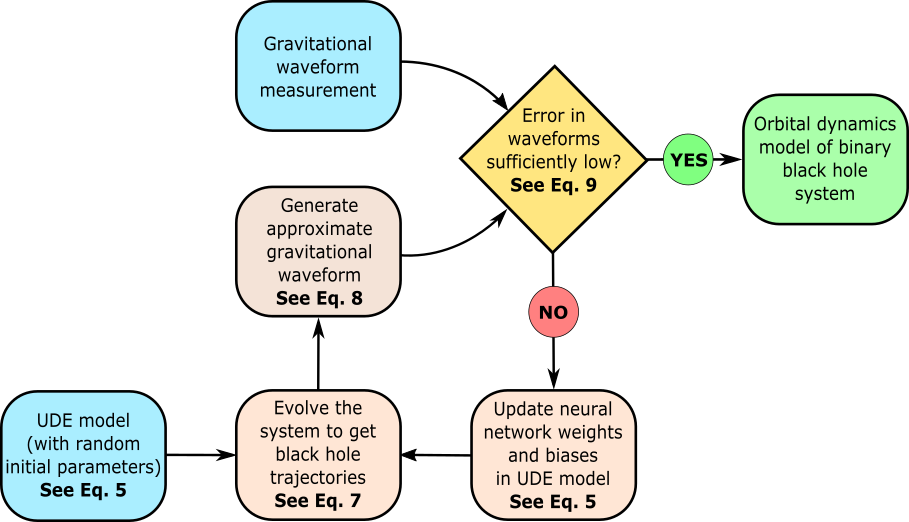}
      \caption{\label{fig:Flowchart}
  Flowchart of algorithm to solve the inverse problem~\cref{eq:InverseProblem}.
  }
\end{figure*}

Upon denoting $\mathbf{x} = (\phi,\chi,p,e)$, we propose the following family of UDEs to describe
the two-body relativistic dynamics:
\begin{subequations}
\label{eq:UDEModel}
\begin{align}
\label{eq:UDEModel_phi}
	\dot \phi
	&=
	\frac{(1+e\cos(\chi))^2}{Mp^{3/2}}
	\big(
		1 + \mathcal{F}_1(\cos(\chi),p,e)
	\big),
	\\\label{eq:UDEModel_chi}
	\dot \chi
	&=
	\frac{(1+e\cos(\chi))^2}{Mp^{3/2}}
	\big(
		1 + \mathcal{F}_2(\cos(\chi),p,e)
	\big),
	\\\label{eq:UDEModel_p}
	\dot p
	&=
	\mathcal{F}_3(p,e),
	\\\label{eq:UDEModel_e}
	\dot e
	&=
	\mathcal{F}_4(p,e),
\end{align}
\end{subequations}
with $\mathbf{x}(0) = (\phi_0,\chi_0,p_0,e_0)$.
Note that the functional form of Eqs.~\eqref{eq:UDEModel} have been inspired by
Eq.~\eqref{eq:EMRIModel}, which are the geodesic equations of motion for an infinitesimally small
``particle" orbiting a super-massive blackhole.
In particular, Eqs.~\cref{eq:UDEModel} are rotationally invariant because the right-hand side omits the $\phi$-variable.
Moreover, when $\mathcal{F}_3 = \mathcal{F}_4 = 0$, orbital energy, $E(p,e)$, and orbital angular momentum, $L(p,e)$, are conserved:
\begin{align}
	\dot{E} = \frac{\partial E}{\partial p}\dot{p} + \frac{\partial E}{\partial e}\dot{e}
	,
	\qquad	\dot{L} = \frac{\partial L}{\partial p}\dot{p} + \frac{\partial L}{\partial e}\dot{e}
	.
\end{align}
Due to the emission of gravitational waves (so-called radiation-reaction), we have that both $\dot{E},\dot{L}<0$ for all time. When each $\mathcal{F}_j = 0$, we recover Newtonian orbits.

Eqs.~\Crefrange{eq:UDEModel_phi}{eq:UDEModel_e} define a family of 
trajectories $\mathbf{x}(t) = \mathbf{x}(t;{\bm{\xi}})$.
Through Eq.~\eqref{eq:body_map}, these trajectories determine the
black hole orbits,
\begin{subequations}
\label{eq:Distance}
\begin{align}
\label{eq:r1}
	\mathbf{r}_1(t)
	=
	\frac{r(t) m_2}{M}\left(\cos(\phi(t)),\sin(\phi(t)), 0 \right) \,,
	\\\label{eq:r2}
	\mathbf{r}_2(t)
	=
	-\frac{r(t) m_1}{M}\left(\cos(\phi(t)),\sin(\phi(t)), 0 \right) 
	\,.
\end{align}
\end{subequations}
Gravitational waves are generated
by orbiting black holes, and so the waves encode detailed information about the 
dynamical variables $\mathbf{r}_1(t)$ and $\mathbf{r}_2(t)$.
General relativity tells us that the dynamics and waves are connected through 
PDEs, which is a familiar scenario in the modeling of waves. 

In the next section
we summarize how to learn $\mathcal{F}_j$ from gravitational wave measurements.
Despite the simplicity of Eqs.~\eqref{eq:UDEModel}, we show that the learned ODEs can describe dynamics beyond the base mechanical model (the base model corresponds to setting the neural network parameters $\bm{\xi}$ to zero and, thus, each $\mathcal{F}_j = 0$). In our first numerical experiment~\ref{sub:extreme_mass_ratio_waveforms}, for example, we show that bound orbits of a test particle following geodesic motion on a Schwarzschild geometry can be accounted for. In our final set of numerical experiments we show that the dissipative dynamics can also be accounted for. Due to the flexible framework of our waveform inversion technique, 
one can easily swap out our base model~\eqref{eq:UDEModel} for others; e.g., the EOB model \cite{buonanno1999effective}.
This suggests many possible future applications of gravitational waveform inversion.

\subsection{Quadrupole formula, the loss function, and model discovery} \label{sec:model_discovery}

Very far from a BBH system, where gravitational wave detectors are located, the gravitational radiation 
field is an outgoing spherical wavefront. On a sufficiently large sphere we can expand the radiation field
into a complete basis of (tensorial) spherical harmonics labeled by $(\ell, m)$ harmonic indices. 

In this paper, we consider only the dominant $(\ell, m)=(2,2)$-mode gravitational waveforms (cf.~\Cref{sub:gravitational_waves_from_an_orbit}), however, our waveform inversion technique could be modified to easily include subdominant modes.
Accordingly, we denote all waveforms by the variable $w = (r/M)\cdot\mathrm{Re}\{h^{22}\}$, where
\begin{align}
\label{eq:gw_22_v2}
h^{22}(t) &=\dfrac{1}{r}\sqrt{\dfrac{4\pi}{5}}\left(\ddot{\mathcal{I}}_{xx}
- 2 i\ddot{\mathcal{I}}_{xy}-\ddot{\mathcal{I}}_{yy}\right)\,
\end{align}
and the trace-free mass quadrupole tensors $\mathcal{I}_{xx}$, $\mathcal{I}_{xy}$, and $\mathcal{I}_{yy}$ are defined in Eq.~\cref{eq:gw_mathcalI} (see also \cite[Eqs.~54--56]{bishop2016extraction}). Eq.~\eqref{eq:gw_22_v2} is the well-known quadrupole formula which 
expresses the measurable waveform, $h^{22}$, in terms of the orbits $\mathbf{r}_1 = (x_1,y_1,0)$ and $\mathbf{r}_2 = (x_2,y_2,0)$. The quadrupole formula
is a very simple approximation that will necessarily introduce systematic error when learning $\mathcal{F}_j$, but 
is sufficient for our purpose. 

We assume that our waveform measurements appear as ordered pairs $(t_k,w_k)$, where $w_k$ denotes the value of the waveform data at time $t_k \in [0,T]$.
In this setting, we define the mean-squared waveform error
\begin{equation}
\label{eq:Loss}
	\mathcal{J}(\mathbf{x})
	=
	\langle J(\mathbf{x},\cdot) \rangle
						:=
	\frac{1}{T}
	\int_{0}^{T}
			J(\mathbf{x},t)
	\,
	\mathrm{d}t
	,
\end{equation}
where $J(\mathbf{x},t) = \sum_{k} \big( w_k - w(t) \big)^2\delta(t-t_k)$
and bracket notation, $\langle \cdot \rangle$, denotes 
denotes averaging over the time interval.
Accordingly, we choose to solve the inverse problem:
\begin{gather}
\label{eq:InverseProblem}
	\min_{{\bm{\xi}}}
	\mathcal{J}(\mathbf{x})
						\quad
	\text{subject to~\Crefrange{eq:UDEModel_phi}{eq:UDEModel_e}}.
\end{gather}
In some situations, convergence to the solution of~\cref{eq:InverseProblem} can be improved by adding well-chosen,
physics-informed penalty and regularization to~\cref{eq:Loss}; cf. \Cref{sub:radiation_reaction_in_numerical_relativity_experiments}.

We note that the exclusive use of gravitational-wave data 
in the loss function is motivated by the consideration 
that in experimental settings only gravitational-wave
observations will be available and never a direct view 
of black hole orbits. Even in computational relativity simulations,
the numerical measurement of black hole trajectories are complicated 
by coordinate ambiguities of general relativity that
make it difficult to assign physical significance to their values.
Waveforms computed from computational relativity simulations,
on the other hand, are well-defined and physically meaningful.

The ODE-constrained optimization problem~\cref{eq:InverseProblem} delivers the calibrated dynamical system model $\dot{\mathbf{x}} = \mathbf{f}(\mathbf{x},\mathcal{F}(\mathbf{x};\bm{\xi}^\star))$, where $\bm{\xi}^\star$ denotes the optimizer found by solving~\cref{eq:InverseProblem}.
This inverse problem can be solved with a number of standard methods.
We choose to use a BFGS algorithm with backtracking line search \cite{nocedal2006numerical} and an adjoint-based (implicit differentiation/adjoint sensitivity method) calculation of gradients \cite{boltyanskiy1962mathematical} implemented with the Julia \cite{bezanson2017julia} software package DiffEqFlux \cite{rackauckas2020universal}.
The algorithm is described by the flowchart in~\Cref{fig:Flowchart}.
Our code is available for download at \cite{JuliaCode}.

\section{Results} \label{sec:examples}

In this section, we present results with three different examples.
The first demonstrates the ability of~\cref{eq:InverseProblem} to recover known orbital equations.
The second two showcase the discovery of new equations of motion for equal mass binary black hole mergers.

\begin{figure*}[!htp]
\centering
	\begin{minipage}{0.55\linewidth}
	  \includegraphics[width=\linewidth]{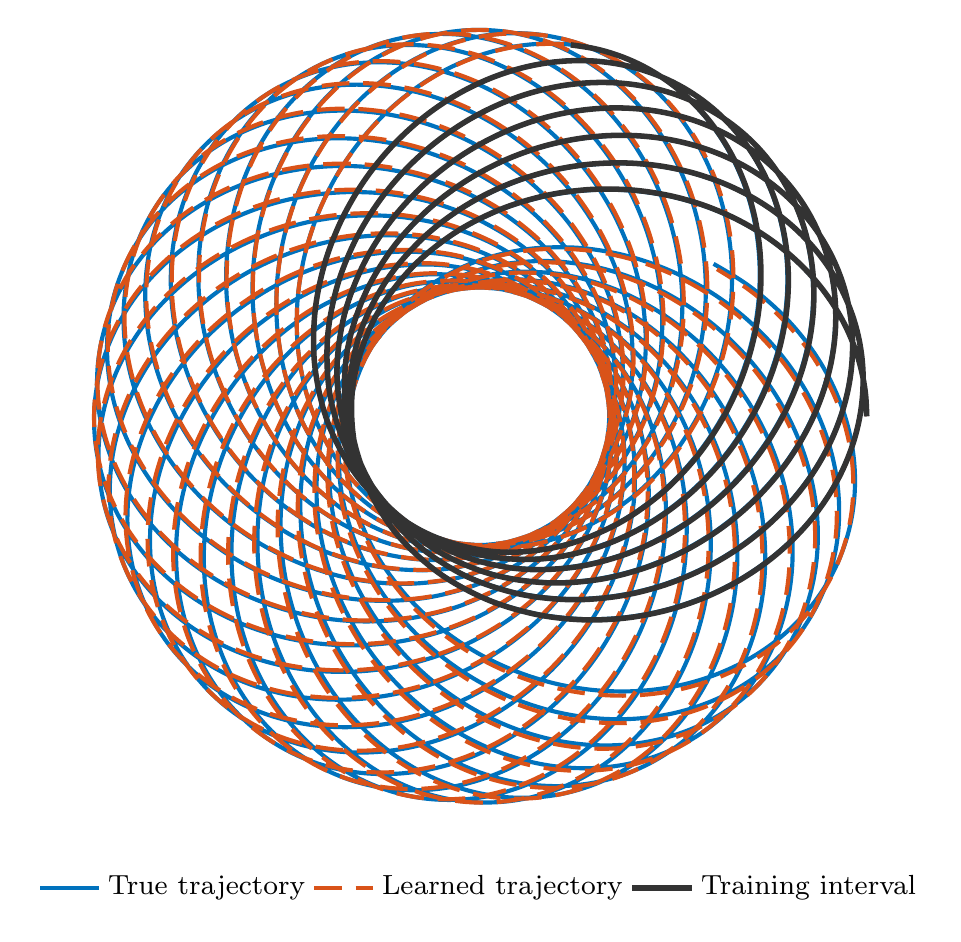}
	\end{minipage}
	~~
	\begin{minipage}{0.32\linewidth}
																						\begin{flushleft}
									\includegraphics[width=\textwidth]{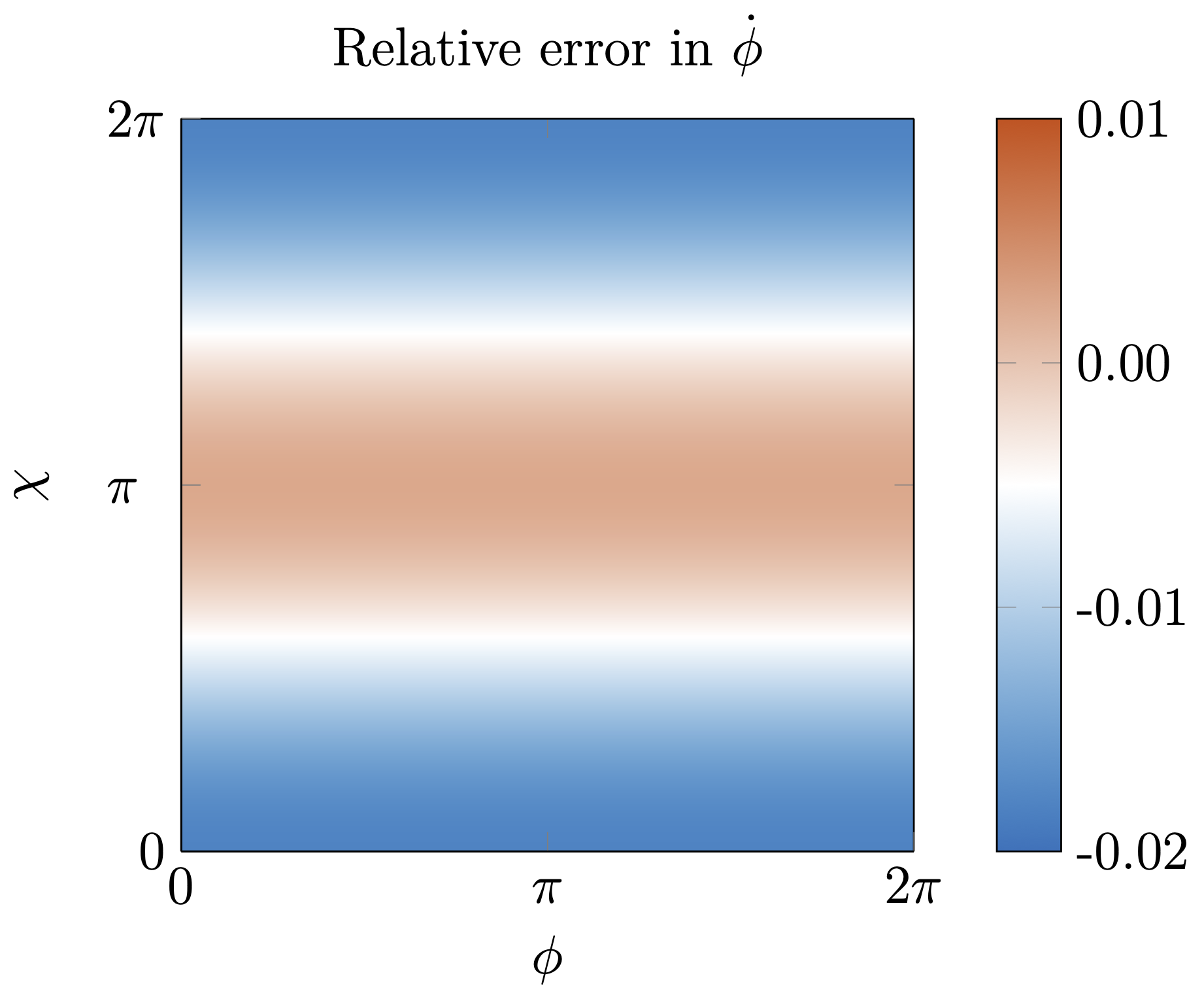}\\[2pt]
			\includegraphics[width=\textwidth]{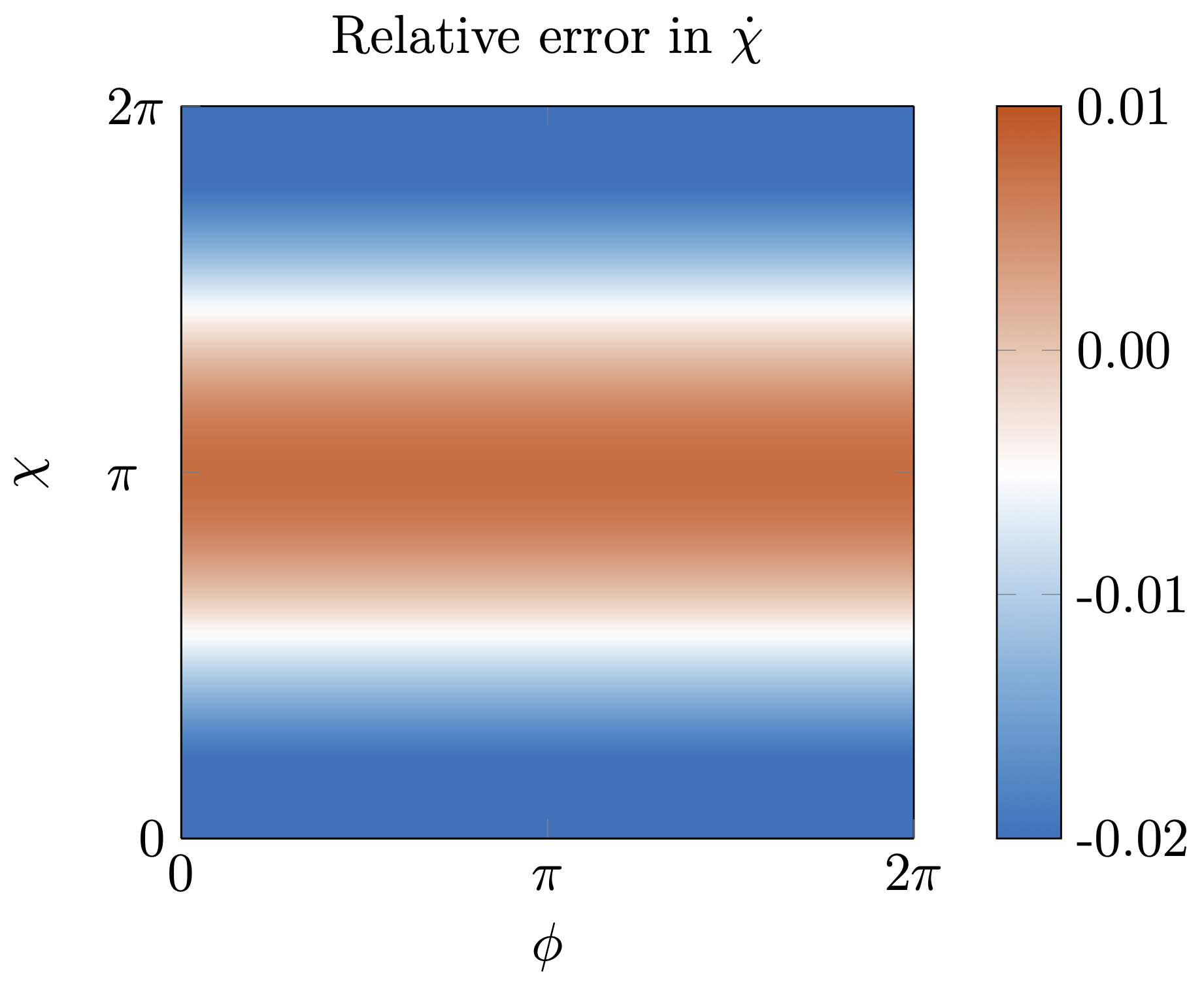}
								\end{flushleft}
	\end{minipage}
	\\
  \includegraphics[width=\linewidth]{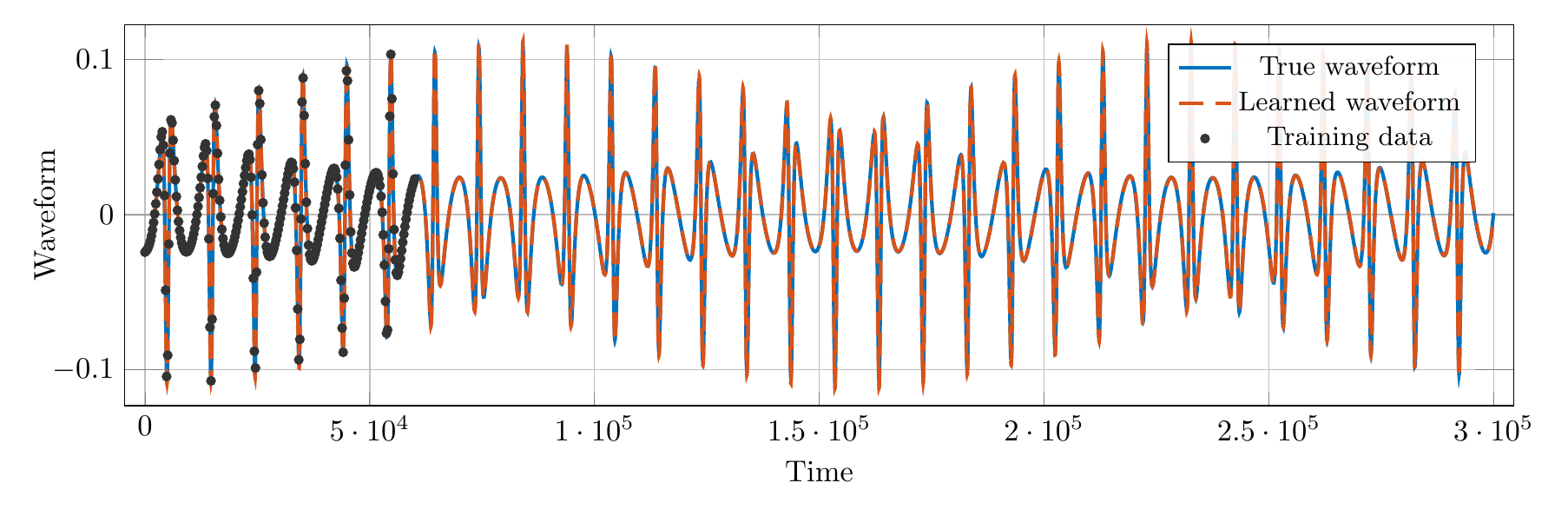}
  \caption{\label{fig:EMRIWaveform} Summary of our first experiment, where we have used gravitational-wave observations
  (black dots; bottom panel) to learn the underlying 
  two-dimensional dynamical system model governing the relativistic two-body problem in the extreme mass ratio limit
  with orbital parameters $p=100,~e=0.5$.
  Top left: Learned (dashed red) and exact (solid blue) trajectories extrapolated $4\times$ the training interval.
  We also show the portion of the orbit (black) corresponding to the gravitational-wave training window, although no
  orbital data was used to learn the dynamics. 
  Top right: Relative error between the learned model~\cref{eq:UDEModel} and the exact model~\cref{eq:EMRIModel}.
  Bottom: Learned (dashed red) and exact (solid blue) waveforms extrapolated $4\times$ the training interval.
  }
\end{figure*}

\subsection{Extreme mass ratio systems} \label{sub:extreme_mass_ratio_waveforms}

As our first motivating example, we consider a special case of the relativistic
two-body problem where the exact solution is known. We show that from short-duration gravitational 
wave observations we are able to discover 
differential equations that are valid over much longer time-scales.

In the regime of $m_1 \gg m_2$, formally the limit $m_1 \rightarrow M$, $m_2$ is a ``test particle"
whose motion obeys~\cite{chandrasekhar1998mathematical,cutler1994gravitational,martel2004gravitational,field2009discontinuous}
\begin{subequations}
\label{eq:EMRIModel}
\begin{align}
	\dot{\phi}
		&=
	\frac{(p-2-2e\cos\chi)(1+e\cos\chi)^2}{Mp^{3/2}\big[(p-2)^2-4e^2\big]^{1/2}}
	,
	\\
	\dot{\chi}
		&=
	\frac{(p-2-2e\cos\chi)(1+e\cos\chi)^2\big[p-6-2e\cos\chi\big]^{1/2}}{Mp^2 \big[(p-2)^2-4e^2\big]^{1/2}}
	,
\end{align}
\end{subequations}
while $\mathbf{r}_1 = (0,0,0)$ and $\dot{e}=\dot{p}=0$.
We shall be interested in the
parameter restriction $0\leq e <1$, for which the radial motion occurs between two
turning points, $pM/(1+e)$ and $pM/(1-e)$ and the orbit is bounded.
When $e = 0$, the orbit is circular.
We let $\mathcal{F}_3 = \mathcal{F}_4 = 0$ and 
provide values for the initial conditions $\phi_0 = 0$ and $\chi_0 = \pi$.
In our example, we set $e = 0.5$, $p = 100$, and $m_1 = 1$, although the results we show 
remain largely the same for other parameter values we have tested. For simplicity, in this
first example we provide known values for $e_0, p_0$ while 
in Sec.~\ref{sub:radiation_reaction_in_numerical_relativity_experiments} we show how
our approach performs when these parameters are also learned. 

To prepare our ground-truth data, 
we numerically solve Eq.~\eqref{eq:EMRIModel} on a dense time grid, thereby 
generating the black hole trajectory $\mathbf{r}_2(t)$. 
We then apply the quadropole formula Eq.~\eqref{eq:gw_22_v2} to generate a gravitational waveform
sampled at $250$ equally-spaced points spanning the time interval $[0, 0.6\cdot 10^5]$,
and shown in Fig.~\ref{fig:EMRIWaveform} (bottom panel; black dots).
Note that in the extreme mass ratio limit,
$m_2 \rightarrow 0$, and the waveform $h^{22} \propto m_2 / m_1$
goes to zero.
Therefore, in this example, we use 
$w = (m_1/m_2) \cdot \mathrm{Re}\{rh^{22}\}$ as gravitational-wave data;
$w$ is now independent of $m_2$.

Using the procedure summarized in Sec.~\ref{sec:model_discovery},
we recover the governing equations by optimizing for $\mathcal{F}_1$ and $\mathcal{F}_2$.
In this setting, both abstract functions only depend on $\cos\chi$.
We exploit this periodic structure by defining $\mathcal{F}_1$ and $\mathcal{F}_2$ with cosine activation functions, $\sigma_j = \cos$.
We then construct $\mathcal{F}_1$ and $\mathcal{F}_2$ as feed-forward neural networks with two hidden layers each; see, e.g.,~Eq.~\cref{eq:TwoHiddenLayers}.
The exact network architecture and numerical discretization can be found in the file \texttt{EMR.jl} in \cite{JuliaCode}.
Finally, we learn the corresponding neural network weights and biases by optimizing~\cref{eq:InverseProblem}.

This process delivers the red trajectory and waveform presented in~\Cref{fig:EMRIWaveform}.
Not only do both waveforms and trajectories match over the training interval consisting of about about 6 orbits, they continue to agree when the learned dynamics are extrapolated to about 31 orbits, after which the orbit's perihelion precession
has undergone a full cycle. 
In \Cref{fig:EMRIWaveform} we compare the true waveforms/trajectories to the learned waveforms/trajectories over the extended time interval $[0, 3\cdot 10^5]$.
To compare the learned model~\cref{eq:UDEModel} to the exact model~\cref{eq:EMRIModel}, we also compute the error in $\dot{\phi}$ and $\dot{\chi}$ over the extended time interval.
Evidently, not only do the waveforms and trajectories match upon visual inspection, the learned model matches the
true mechanical model to about two orders of magnitude. The learned model also recovers important 
relativistic effects, notably perihelion precession, from just a few gravitational-wave cycles. Finally, we note 
that once the dynamical model is known, it can be used to generate very long orbits and gravitational wave
signals by integrating the ODE~\cref{eq:UDEModel} and post-processing the solution with the quadrupole formula~\cref{eq:gw_22_v2}.

This experiment demonstrates the potential power of waveform inversion in a simple setting with a known
solution.
It also demonstrates how the information content in the original waveform can be used to infer UDEs.
Nevertheless, this system is conservative ($\dot{e}=\dot{p}=0$), the quadrupole formula is prescribed exactly, and the learnable dynamics depend only on the $\chi$-variable. The following examples are more challenging because none of the aforementioned simplifications hold.

\begin{figure*}
\centering
	\begin{minipage}{0.45\linewidth}
		\centering
		\includegraphics[width=\linewidth]{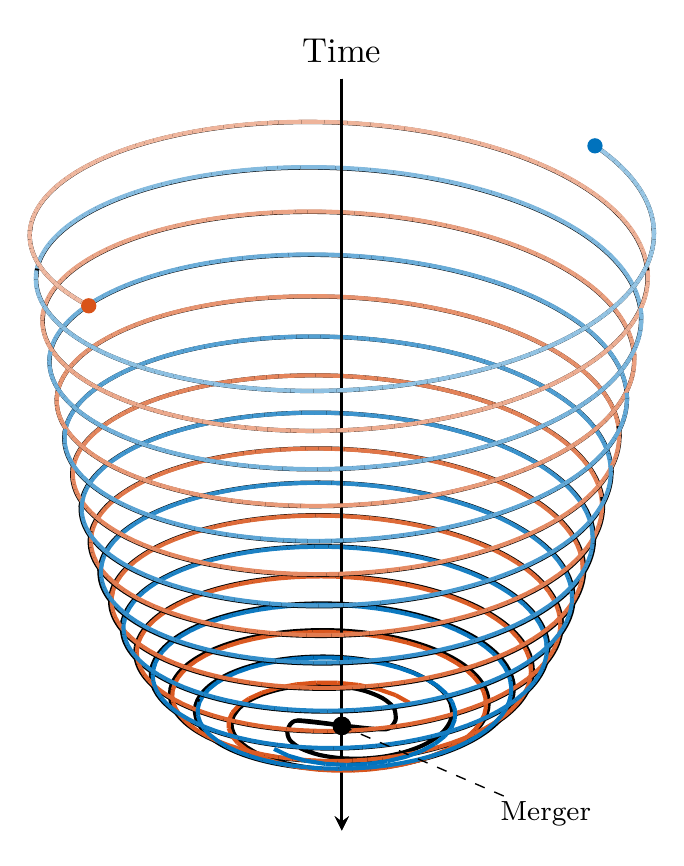}
			\end{minipage}
																						\begin{minipage}[t]{0.5\linewidth}
					  \includegraphics[height=4.5cm]{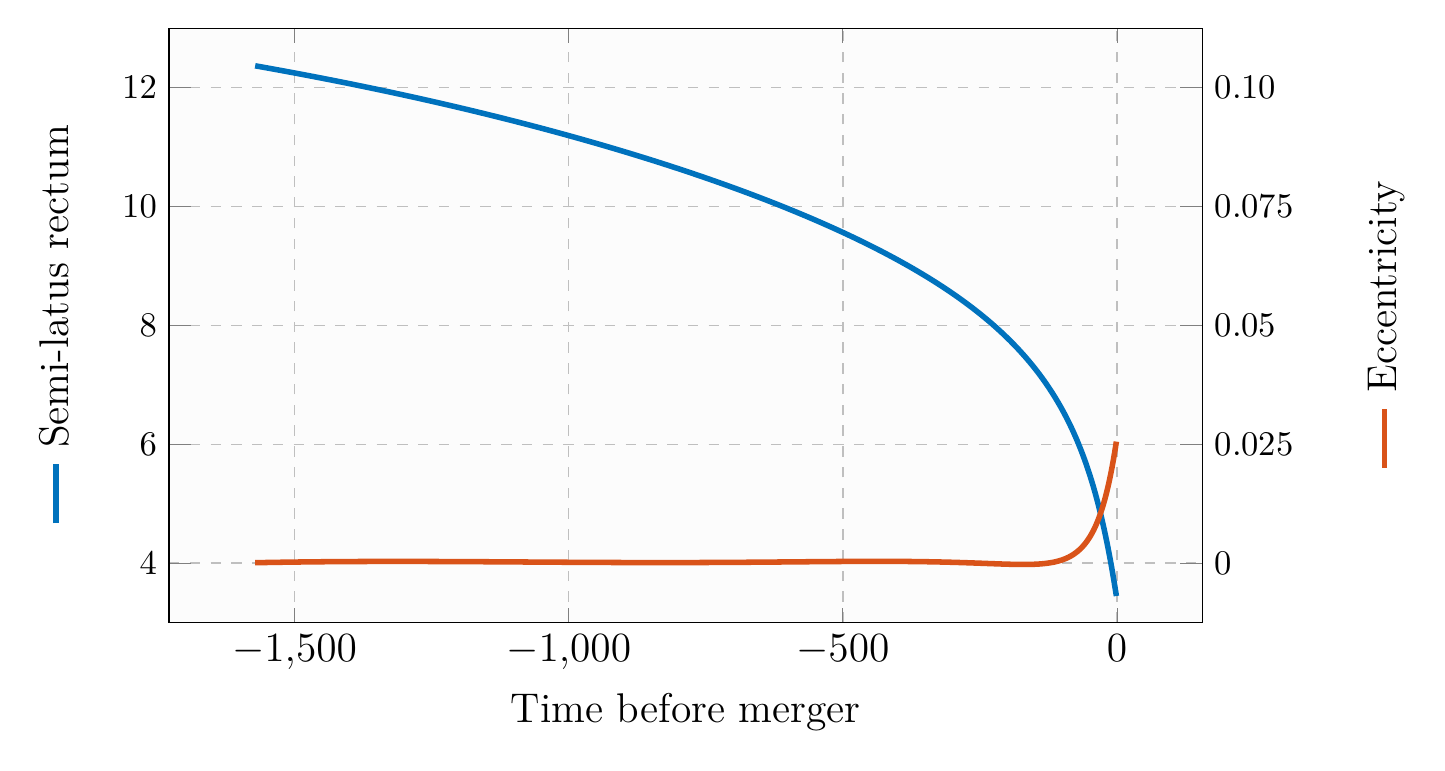}\\[4pt]
	  \includegraphics[height=4.5cm]{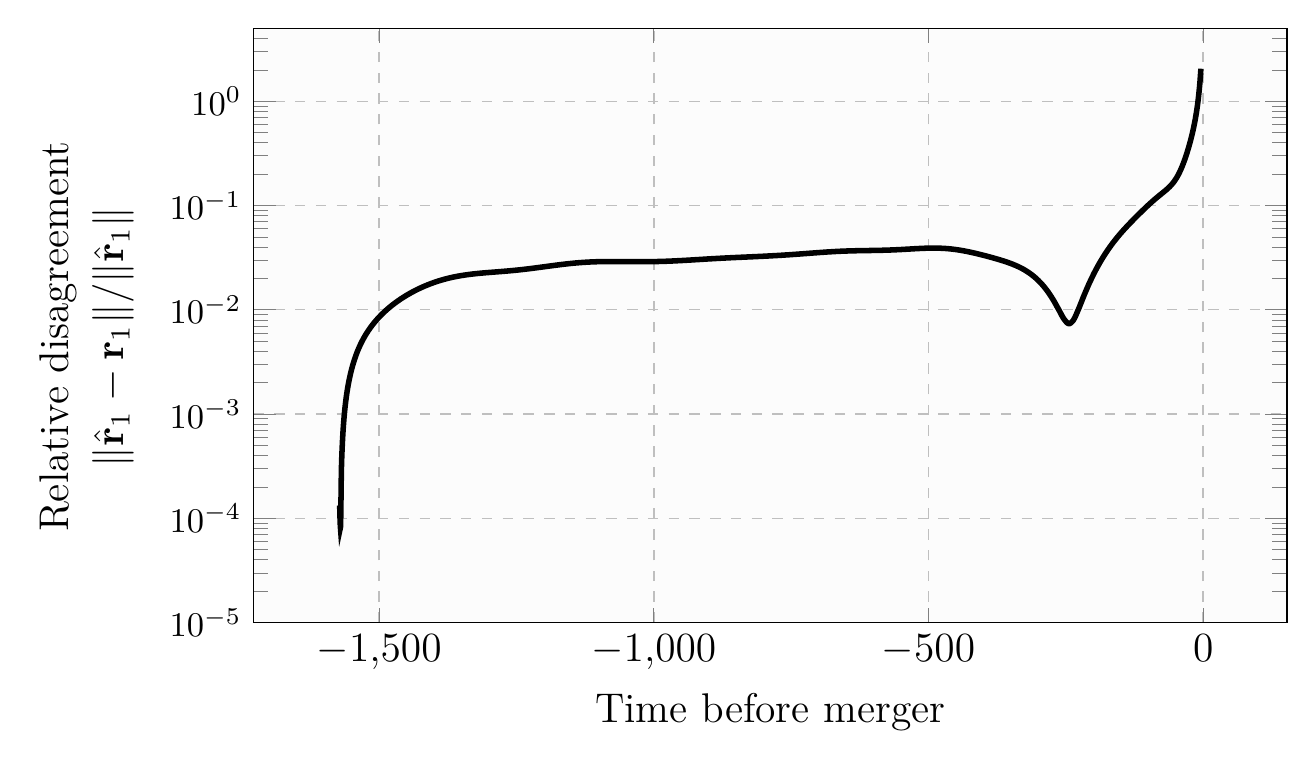}
	\end{minipage}
  \includegraphics[width=0.95\linewidth]{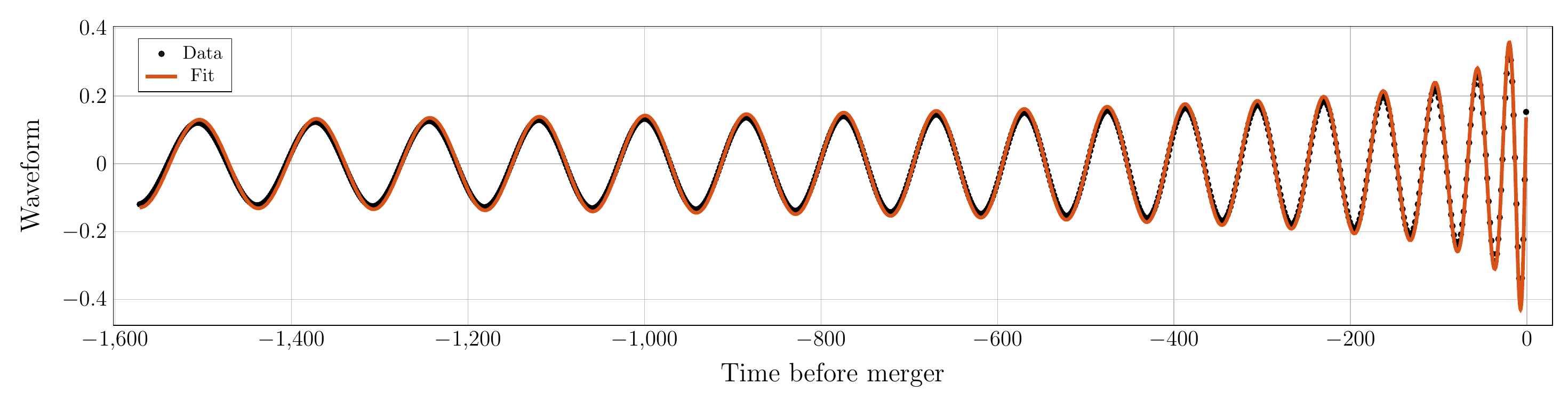}
  \caption{\label{fig:NR1Trajectories} Summary of our second experiment, where have used gravitational-wave observations
  (black dots; bottom figure) to learn the underlying dynamical system model governing the relativistic two-body problem for two equal mass black holes in quasi-circular orbit. 
  Top left: Trajectories of the centers of mass of the black hole system \texttt{SXS:BBH:0217} (black lines), taken from the SXS Gravitational Waveform Database \cite{Boyle:2019kee}. We also show the orbit computed from our learned dynamical system (blue and red lines). In the upper right panel, 
  we show the evolution of the eccentricity and semi-latus rectum from our learned-dynamical system. The middle right panel shows the disagreement between the NR trajectories and the ones computed from the learned dynamical system. We caution the reader that this figure should not be understood as a relative error because the numerical relativity black hole trajectories and our learned model are expressed in different coordinate systems that are impossible to relate. 
  Bottom: Learned (red line) and computational relativity (black dots) waveform data corresponding to the real part of the $h^{22}$ mode.
      }
\end{figure*}

\subsection{General relativistic orbital dynamics of binary black holes} \label{sub:radiation_reaction_in_numerical_relativity_experiments}

In this pair of examples, we consider numerically generated waveform measurements from equal mass $m_1 = m_2 = 0.5$
binary black hole systems. Unlike the previous experiment, the orbital dynamics for these systems 
is much more complicated; the exact equations of motion are unknown and the dynamics include time-dependent
values of the eccentricity and semi-latus rectum. 
Although an extensive body of literature exists for deriving these equations 
from approximations of general
relativity~\cite{blanchet2014gravitational,buonanno1999effective,poisson2011motion}, we are unaware of any data-driven approaches focused on discovering orbital dynamics from waveform measurements.

Computational relativity codes provide 
exact (up to numerical discretization error) solutions to
the general relativistic two-body problem, including both the 
corresponding trajectories of the center of the black holes and 
gravitational-wave data. 
Although the location of the black holes
are 
coordinate-dependent they can still be used to compare with the trajectories obtained from our model.
However, such comparisons should no longer be understood as model error since the coordinate system
used for the data and model are necessarily different.

The simulations for this work 
were performed using the Spectral Einstein
Code (SpEC)~\cite{SpECwebsite,York:1998hy, Pfeiffer:2002iy, Lindblom:2005qh,Rinne:2008vn,Boyle:2019kee,SXSCatalog} 
developed by the Simulating eXterme Spacetimes (SXS)
collaboration~\cite{SpECwebsite} and made publicly available through the 
Gravitational Waveform Database \cite{Boyle:2019kee}.

From now on, we augment the loss function $\mathcal{J}$ in~\cref{eq:Loss} with non-negative penalty and regularization terms, motivated below:
\begin{equation}
\label{eq:NewJ}
	\mathcal{J}(\mathbf{x},\bm{\xi})
	=
	\langle J(\mathbf{x},\cdot) \rangle
	+
	\mathcal{P}_1(\mathbf{x})
	+
	\mathcal{P}_2(\mathbf{x})
	+
	\mathcal{R}(\bm{\xi})
	.
\end{equation}
In this new expression, we define
\begin{equation}
\label{eq:Penalty}
						\mathcal{P}_1(\mathbf{x})
	=
	\gamma_1\langle (\dot{p})_+^2 \rangle
	+
	\gamma_2\langle (\ddot{p})_+^2 \rangle
			,
\end{equation}
where $(f(t))_+ = \max\{f(t),0\}$,
\begin{equation}
\label{eq:Penalty2}
	\mathcal{P}_2(\mathbf{x})
	=
	\gamma_3\langle (-e)_+^2 \rangle
	+
	\gamma_4\langle (e-e_0)_+^2 \cdot \bm{1}_{\{p>6+2e_0\}} \rangle
	,
\end{equation}
where $\bm{1}_{\Omega}$ denotes the indicator function on the set $\Omega \subset [0,T]$, and finally
\begin{equation}
\label{eq:Regularization}
	\mathcal{R}(\bm{\xi})
	=
	\gamma_5\|\bm{\xi}\|^2
	,
\end{equation}
where $\|\bm{\xi}\|$ denotes the $\ell^2$-norm of the expanded parameter vector $\bm{\xi}$.
It is standard practice to use large coefficients for penalty terms and small coefficients for regularization terms.
However, as explained in the paragraphs below, our penalty terms are present to help avoid nonphysical local minima and are not active in the optimized model. For this reason, we do not make a concerted attempt to tune these coefficients.
In both of the coming experiments, we somewhat arbitrarily fix $\gamma_1 = 10^3,~ \gamma_2 = 10^2,~ \gamma_3 = 10^1$, and $\gamma_5 = 10^{-1}$.
In the first experiment, we take $\gamma_4 = 1$, while in the second experiment we use $\gamma_4 = 0$.

The physical motivation for the terms in~\cref{eq:Penalty} relies on Eqs.~\cref{eq:Distance}.
From these equations, we have that the distance between the two black holes $r$ is proportional to $p$.
Due to energy loss from the emitted gravitational waves, $r(t)$ converges to zero at a rate
that increases throughout the system's evolution. 
The penalty terms $\langle (\dot{p})_+^2 \rangle$ and $\langle (\ddot{p})_+^2 \rangle$ have been chosen to encourage the selection of solutions with this physical behavior.
The first term in Eq.~\cref{eq:Penalty2} encourages the selection of a
positive eccentricity function $e(t)$ for all time $t$.
On the other hand, the final term in this definition is motivated by the stability condition $p \geq 6+2e$ for bound orbits.
It is widely accepted that $e$ decays in this range \cite{cutler1994gravitational}, and this term helps to direct the solution toward models with this property.

Clearly, if $\dot{p},\ddot{p},-e,(e-e_0)^2\cdot \bm{1}_{\{p>6+2e_0\}} \leq 0$, then $\mathcal{P}_1(\mathbf{x}) = \mathcal{P}_2(\mathbf{x}) = 0$.
Our experiments appear to indicate that optimal solutions $p(t)$ and $e(t)$ satisfy each of these bounds, therefore, the penalty terms only act as guardrails throughout the optimization process.
The Tikhonov--Phillips regularization term~\cref{eq:Regularization} helps convergence by ensuring continuous dependence between the data and the solution \cite{benning_burger_2018}.
The Tikhonov regularization term $\|\bm{\xi}\|^2$ can also help avoid model degeneracies and overfitting in the presence of noisy data.
For example, when the orbit is circular ($e=0$)
the model is degenerate in $\chi$ and, therefore, it is also degenerated in the weights and biases defining $\mathcal{F}_2$.
Other penalty and regularization terms could be considered in future studies.

In the following pair of examples, we construct feed-forward neural network parameterizations of $\mathcal{F}_j$, $j=1,\ldots,4$, with $\tanh$ activation functions.
The exact network architecture we use can be found in files \texttt{SXS1.jl} and \texttt{SXS2.jl} \cite{JuliaCode}.

\subsubsection{Near-circular orbits from clean GW observations} \label{ssub:low_eccentricity_orbit}

For this experiment, we consider a binary black hole system with negligible eccentricity during the initial inspiral.
For inspection, the center of mass-corrected trajectories of the binary black hole system are depicted in the top left-hand corner of~\Cref{fig:NR1Trajectories} (solid black lines), with the associated 1000 equally-spaced waveform data points
in the bottom panel (black dots).
From now on, we let $[0,T]$ denote the time interval between the first ($t=0$) and final ($t=T$) measurement, where the final measurement occurs shortly \emph{before} merger.

As in the previous experiment, we adopt the initial conditions $\phi_0 = 0$ and $\chi_0 = \pi$
and assume  that $r_0$ is known.
Using Eq.~\eqref{eq:r_parameterization}, these assumptions provide us with an explicit expression for $p_0$, $M p_0 = r_0\cdot(1+e_0\cos(\chi_0))$.
Due to the nearly-zero eccentricity of the initial trajectories, we opt for the simple initial condition $e_0 = 0$.
In the next and final experiment, we treat the more realistic case where both $e_0$ and $\chi_0$ are unknown.

In order to avoid local minima, we solve~\cref{eq:InverseProblem} on a sequence of increasing time intervals $[0,T_0] \subsetneq [0,T_1] \subsetneq \cdots \subsetneq [0,T]$, using the optimal parameters $\bm{\xi}^\star$ from each preceding optimization problem (plus a small amount of Gaussian noise) as initial data for the subsequent problem.
Using this incremental procedure, we are able to recover the overwhelming majority of the black hole trajectories, as indicated by the visual agreement between the learned (red and blue) and NR (black) trajectories shown in the top left-hand panel of~\Cref{fig:NR1Trajectories}. We also depict the relative disagreement between the NR trajectory of the first black hole $\hat{\mathbf{r}}_1$ and the learned trajectory ${\mathbf{r}}_1$. The model also recovers important general
relativistic effects, notably the learned functions $\mathcal{F}_3$ and $\mathcal{F}_4$ cause a 
runaway inspiral process that drives the black holes to merge.
This process is seen most clearly by monitoring the behavior of $p(t)$
in the upper right-hand panel of~\Cref{fig:NR1Trajectories}.
We also observe that our model is able to naturally include both the inspiral ($p > 6$)
and plunge ($p< 6$) orbital regimes. Note the upper right-hand panel of~\Cref{fig:NR1Trajectories} shows 
that near this transition region the eccentricity quickly grows, which is at odds with 
our physical expectation for stable orbits and indicates a very different dynamical regime.

One complication in validating our learned dynamical system is how to perform meaningful comparisons with other models. Indeed, besides the waveform, the other three sub-panels shown in Fig.~\ref{fig:NR1Trajectories} depict gauge-dependent quantities; that is their value depends on the coordinate system being used. In particular, our trajectories are not expressed in 
the same damped harmonic gauge coordinates used by SpEC simulations~\cite{szilagyi2009simulations}. Nevertheless, recent studies have noted surprisingly good agreement between NR trajectories and those computed with post-Newtonian (PN) models~\cite{boyle2014gravitational} and PN-augmented dynamical models~\cite{varma2019binary,blackman2017surrogate,blackman2017numerical}. In particular, Ref.~\cite{varma2019binary} conjectures that the main source of disagreement is due to the PN formula being expressed in the harmonic gauge~\cite{blanchet2014gravitational}. The close agreement between NR and UDE trajectories shown in Fig.~\ref{fig:NR1Trajectories} is another example of surprisingly good agreement~\cite{varma2019binary,boyle2014gravitational}.

To avoid gauge ambiguities, comparisons of BBH dynamics focus on comparing gauge-invariants that are computable within different frameworks. For example, Refs.~\cite{damour2012energy,nagar2016energetics} explore the conservative dynamics by comparing the relationship between the total energy and total angular momentum from NR data  to the corresponding analytical predictions from PN and EOB theory. However, the identification of a conserved energy or angular momentum within our setup is not obvious as our equations are not derived from a Hamiltonian; we will return this issue in the Discussion section. Instead, we follow Ref.~\cite{van2020intermediate} and compute the accumulated orbital phase as a function of the orbital frequency. This quantity includes both dissipative and conservative effects, and can be computed within different modeling frameworks.

\begin{figure}
	\centering
	\includegraphics[width=\linewidth]{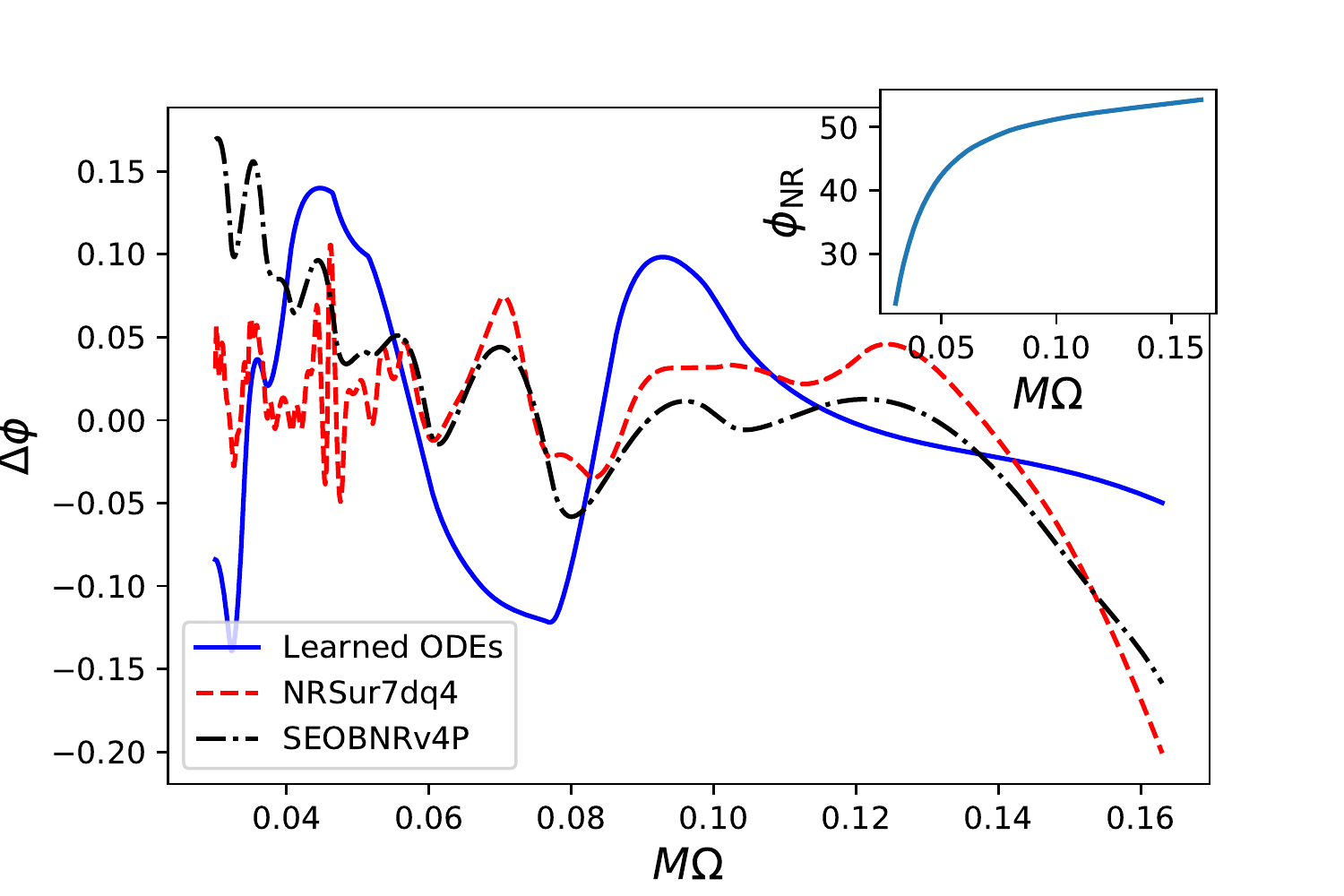}
  \caption{\label{fig:phase_comparison} Comparison of the accumulated orbital phase as a function of orbital frequency for the second experiment~\ref{ssub:low_eccentricity_orbit}, an equal mass quasi-circular inspiral of nonspinning black holes. The orbital phasing obtained from an NR simulation is shown in the inset figure. We show the difference between the NR data and (i) our UDE orbital model (solid blue line), (ii) {\tt NRSur7dq4}, a numerical relativity surrogate model (dashed red line), and (iii) {\tt SEOBNRv4PHM}, a recently developed precessing EOB model (dash-dot black line). All three models show good agreement with the NR orbital phasing. The UDE (learned ODE) model is competitive with these two state-of-the-art models despite being built in a very different way.}
\end{figure}

The NR orbital phase is defined in terms of the waveform data as follows:
\begin{equation}
\label{eq:nr_orbital_phase}
\phi_{\rm NR}^{\rm orb}(t) =\frac{1}{2} \arg h_{22}(t) \,.
\end{equation}
We also compute the orbital phase from a recently developed precessing
EOB model ({\tt SEOBNRv4PHM})~\cite{ossokine2020multipolar}, a numerical relativity surrogate model ({\tt NRSur7dq4})~\cite{varma2019surrogate}, and our UDE model. Both {\tt NRSur7dq4} and {\tt SEOBNRv4PHM} are considered state-of-the-art and have been used by the LIGO-Virgo Collaboration to analyze recent gravitational-wave observations~\cite{abbott2020gw190521,abbott2020gw190412,abbott2020gwtc}. We represent each orbital phase by a degree 3 spline (using a smoothing factor of 0.0002), from which we compute the orbital frequency,
\begin{equation}
\label{eq:orbital_frequency}
\Omega = \frac{d \phi^{\rm orb}}{dt} \,,
\end{equation}
by taking a derivative of the spline, and finally forming the function $\phi^{\rm orb}(\Omega) = \phi^{\rm orb}(t(\Omega))$. We then compare $\phi_{\rm NRSur7dq4}^{\rm orb}$, $\phi_{\rm SEOBNRv4P}^{\rm orb}$, and $\phi_{\rm UDE}^{\rm orb}$ to $\phi_{\rm NR}^{\rm orb}$. For each comparison, we form the difference, $\Delta \phi = \phi_{\rm model}^{\rm orb} - \phi_{\rm NR}^{\rm orb}$ after phase alignment\footnote {We allow for $\phi_{\rm model} \rightarrow \phi_{\rm model} + c$, for some constant $c$, which is equivalent to a rotation in the orbital plane.}. Figure \ref{fig:phase_comparison} shows $\Delta \phi$ for each model. Over the range of orbital frequencies shown, the $L_2$-error in the phase is $2.5\times 10^{-2}$ (UDE model), $2.0\times 10^{-2}$ ({\tt NRSur7dq4}), and $2.2 \times 10^{-2}$ ({\tt SEOBNRv4PHM}). All three models do an excellent job at tracking the NR orbital phase throughout the late inspiral phase to $\Omega \approx 0.16$, which is when a common apparent horizon appears in the NR simulation.

One of the most important practical uses of a dynamical model is as an intermediate step towards generating gravitational waveforms. While a full study is outside the scope of this paper, we provide a preliminary look at this here. We compute an $L_2$-type error measurement (see Eq.~21 from Ref.~\cite{blackman2017surrogate}) between the complexified NR waveform and each model's prediction of the waveform after optimizing for phase and time alignments. We find the errors to be $3.1 \times 10^{-3}$ (UDE model), $1.2 \times 10^{-3}$ ({\tt SEOBNRv4PHM}), and $1.1 \times 10^{-5}$ ({\tt NRSur7dq4}). All three models have been calibrated to $q=1$ NR waveform data, so this comparison is only meant to be suggestive of how well the modeling techniques can perform on the training set and not its generalization error.

We note that all three models used in our comparisons have been built in very different ways. The {\tt SEOBNRv4PHM} model is a highly sophisticated analytical model thats been under investigation for two decades~\cite{buonanno1999effective} while the {\tt NRSur7dq4} model was trained against 1528 NR simulations using numerical techniques that have been in development for nearly a decade~\cite{field2011reduced,field2014fast}. By comparison, our UDE model is new and our modelization choices are simple. Given that our UDE model is able to perform comparably well against state-of-the-art models demonstrates the potential of waveform inversion as new tool for model builders to consider in future work.

\subsubsection{Eccentric orbits from noisy GW measurements} \label{ssub:low_eccentricity_orbit}

\begin{figure*}
\centering
	\begin{minipage}{0.45\linewidth}
	\centering
		\includegraphics[width=\linewidth]{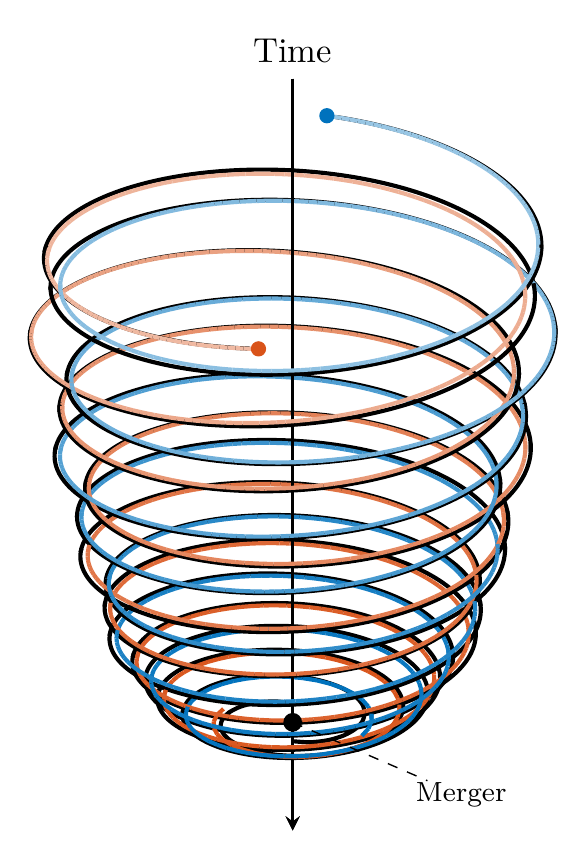}
			\end{minipage}
		\begin{minipage}[t]{0.5\linewidth}
					\includegraphics[height=4.5cm]{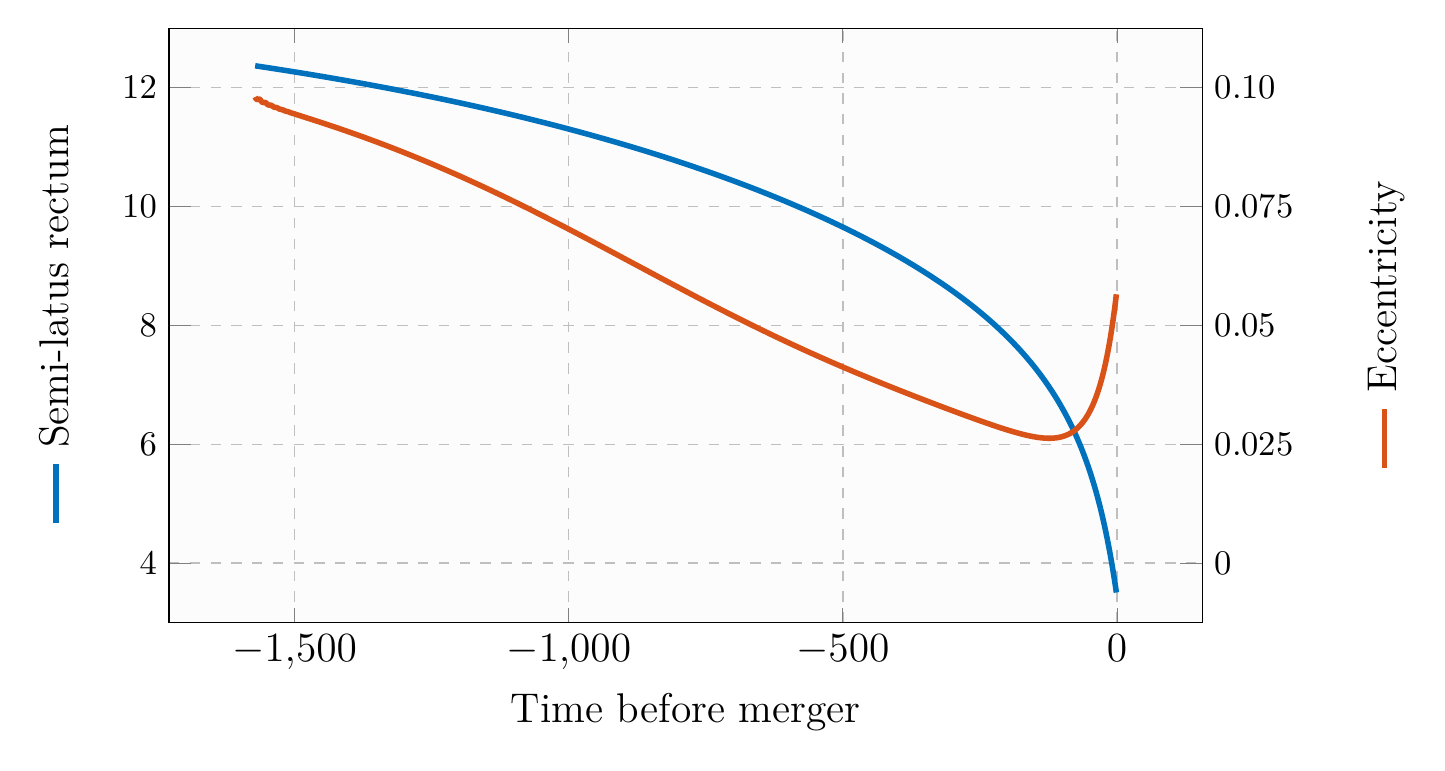}\\[11pt]
		\includegraphics[height=4.5cm]{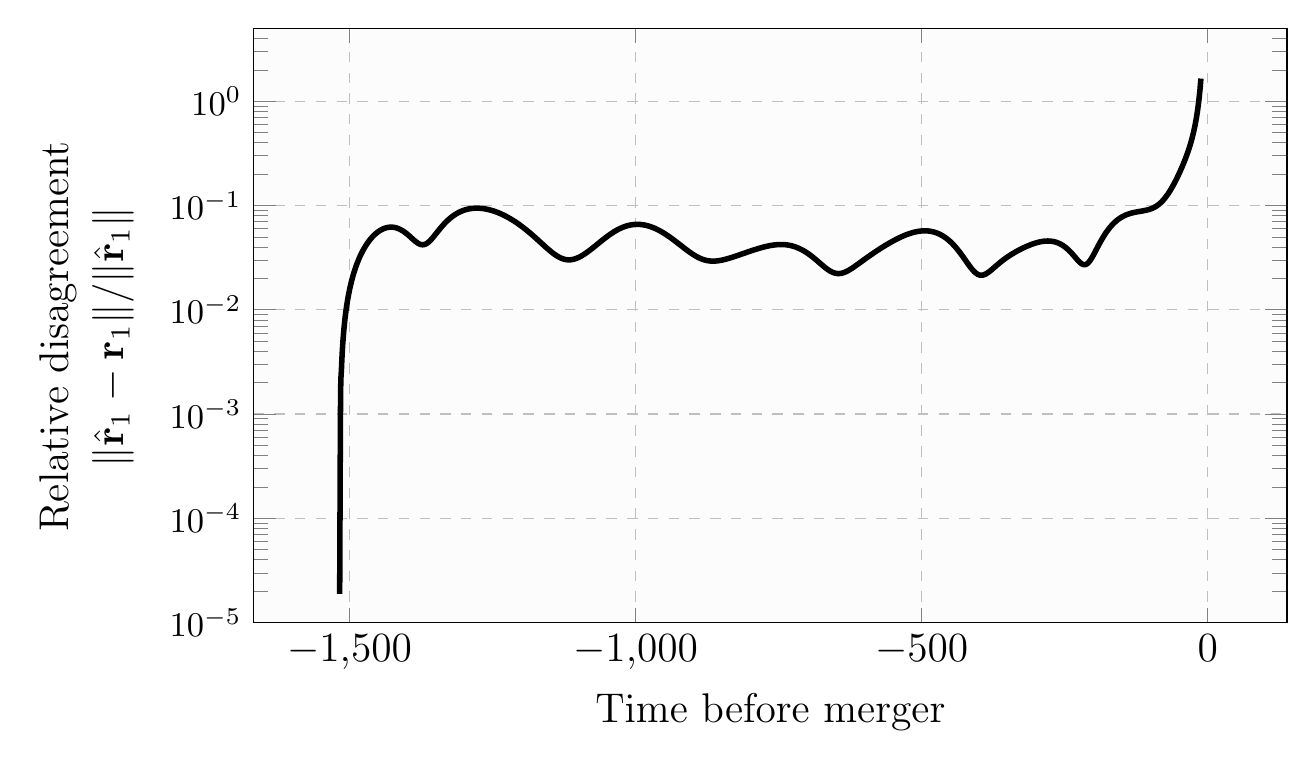}
	\end{minipage}
  \includegraphics[width=0.95\linewidth]{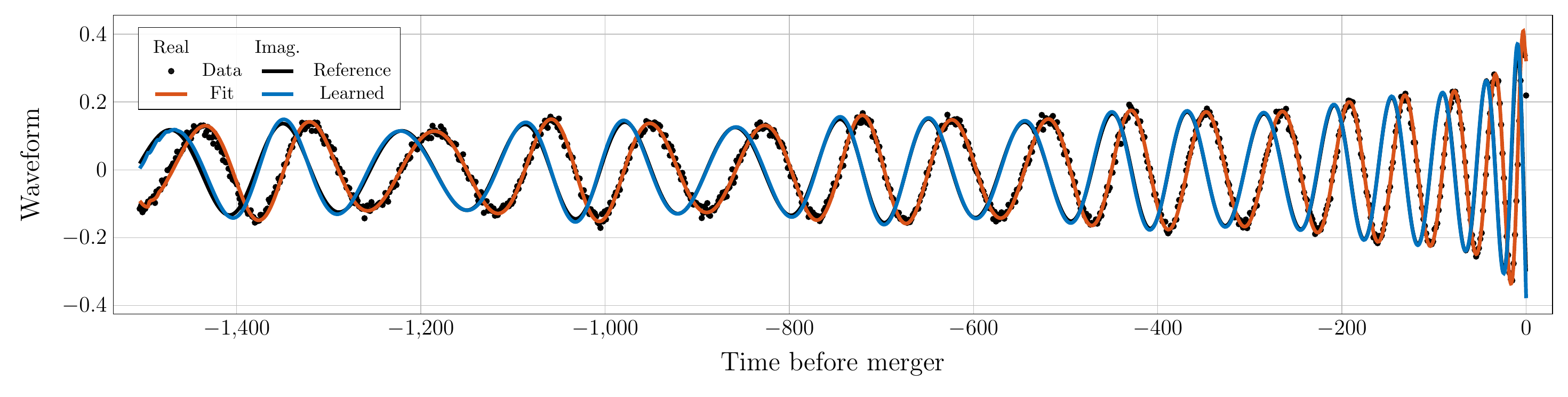}
  \caption{\label{fig:NR2Trajectories}
  Summary of our third experiment, where have used noisy gravitational-wave observations
  (black dots; bottom figure) to learn the underlying dynamical system model governing the relativistic two-body problem for two equal mass black holes in an eccentric orbit. 
  Top left: Trajectories of the centers of mass of a black hole system \texttt{SXS:BBH:1356} (black lines), taken from the SXS Gravitational Waveform Database \cite{Boyle:2019kee}. We also show the orbit computed from our learned dynamical system (blue and red lines). In the upper right panel, 
  we show the evolution of the eccentricity and semi-latus rectum from our learned-dynamical system. The middle right panel shows the disagreement between the NR trajectories and the ones computed from the learned dynamical system. We caution the reader that this figure should not be understood as a relative error because the numerical relativity black hole trajectories and our learned model are expressed in different coordinate systems that are impossible to relate. 
  Bottom: Learned (red line) and computational relativity (black dots) waveform data corresponding to the real part of gravitational waveforms.
  Here, we also include the learned imaginary part of the waveform (blue line), reconstructed with the quadrupole formula, and compare it with the reference imaginary part taken from the SXS database (black line).
   }
\end{figure*}

This experiment proceeds in much the same way as the previous one.
Here, however, we learn the dynamics of an eccentric binary black hole 
system whose trajectories are depicted in the top left-hand corner of~\Cref{fig:NR2Trajectories} (solid black lines) with the associated waveform data in the bottom panel (black dots).
Unlike the previous experiments, we do not assume known values for the initial conditions $e_0,p_0,$ or $\chi_0$ but instead 
make these part of the learning process. We continue to adopt the initial conditions $\phi_0 = 0$.
As can be seen in \Cref{fig:NR2Trajectories}, we introduce additive Gaussian noise to the waveform data of the form
$w(t_i) + n(t_i)$, where $n(t_i)$ is draw from a normal distribution of mean 0 and standard deviation of $\sigma = 10^{-2}$. 
As the typical waveform amplitude is $\sim 0.1$, this corresponds to a coefficient of variation of around $\sigma/0.1 = 0.1$.

In spite of these new challenges, we are still able to recover 
the original trajectories 
as indicated by the visual agreement between the learned (red and blue) and NR (black) trajectories shown in the top left-hand panel of~\Cref{fig:NR2Trajectories}.
This is achieved by simultaneously optimizing for both $e_0$ or $\chi_0$ in~\cref{eq:InverseProblem}, in addition to the neural network parameters $\bm{\xi}$, and deducing the associated value of $p_0$ directly from Eq.~\cref{eq:r_parameterization}.
The model also recovers important general
relativistic effects, notably, as in the previous example, 
radiation-reaction effects. In this case, 
the tendency for the orbit to circularize, $e(t) \rightarrow 0$, is seen
in the upper right-hand panel of~\Cref{fig:NR2Trajectories}. As before, the 
eccentricity has an inflection point around $p \approx 6$ and quickly grows thereafter.

\begin{figure}
\begin{center}
		\includegraphics[width=\linewidth]{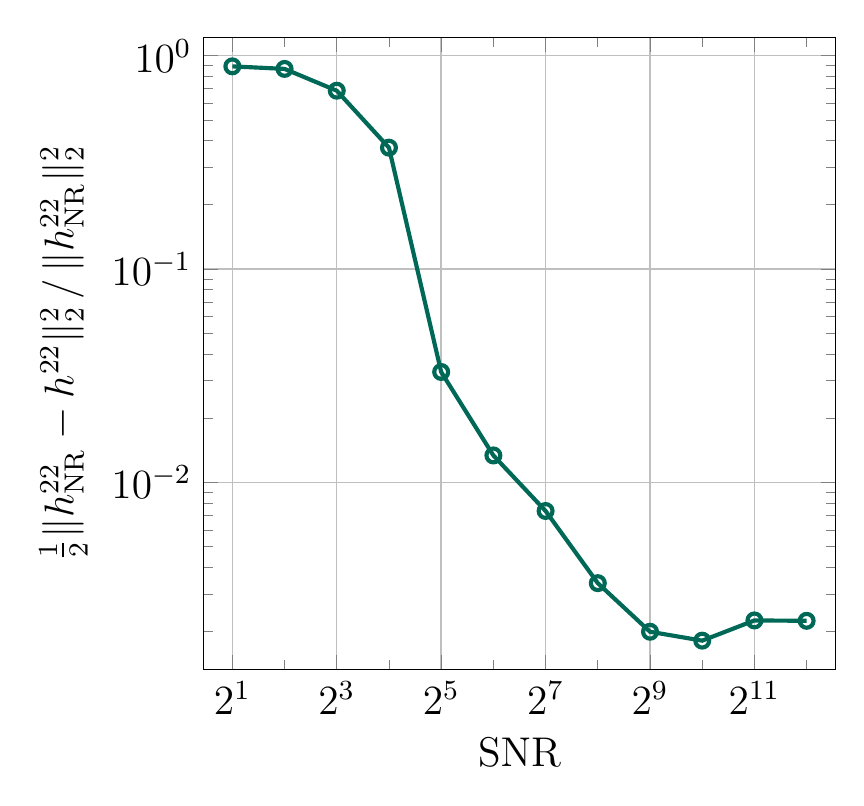}
\end{center}
  \caption{\label{fig:vsSNR}
  We quantify the method's robustness to increasingly amounts of Gaussian 
  noise by comparing the true NR waveform, $h_{\rm NR}^{22}$, to the learned waveform, $h^{22}$. The amount of noise is quantified by the signal's SNR. We compute the waveform error according to $\frac{1}{2}\|h_{\rm NR}^{22} - h^{22}\|_2^2\, /\, \|h_{\rm NR}^{22}\|_2^2$, which is related to the overlap error (cf. Ref.~\cite{blackman2017surrogate}) commonly used in gravitational-wave data analysis.
  } 
\end{figure}

As follow-on to this experiment, we test the stability of the learned solution to different signal-to-noise ratios (SNRs).
Specifically, we use the learned solution parameters $\bm{\xi}$, $e_0$, and $\chi_0$ from~\Cref{fig:NR2Trajectories} as initial guesses in a new set of model discovery problems that fit the original \texttt{SXS:BBH:1356} waveform data, but have white Gaussian noise of different variance superimposed. After solving each of these new model discovery problems, we measure how much the learned waveform differs from the true waveform. The curve in~\Cref{fig:vsSNR} shows the relative error in the learned waveform as a function of the SNR in the waveform measurement. The relative error grows as the SNR decreases, however, the results are surprisingly accurate down to SNR$\approx 32$. We note that while all BBH gravitational-wave detections to-date have had SNRs below 32~\cite{abbott2020gwtc}, future detectors such as LISA~\cite{2017arXiv170200786A}, Cosmic Explorer~\cite{Evans:2016mbw}, or Einstein Telescope~\cite{Hild:2010id} will
routinely detect events with an SNR greater than 32.

\section{Discussion} \label{sec:discussion}
 
We have presented a new, data-driven
gravitational waveform inversion strategy which generates
mechanical models of binary black hole systems. We start with a structurally very simple set of universal differential equations
and parameterize the space of models with feed-forward neural networks. Our differential equations are trained 
by solving a physics-informed constrained optimization problem that seeks to minimize the waveform error. We tested our method 
on various BBH systems including extreme and comparable mass ratio systems in eccentric and non-eccentric orbits, 
and train on portions of the waveform corresponding to orbital plunge right up to the time of merger.
We find that the resulting differential equations agree remarkably well 
with the black hole trajectories computed through purely numerical means.
Our models can be extrapolated in time and recover various known relativistic effects despite these being 
previously unknown to the universal differential equations. The main contribution of our work is to show that
two-body relativistic models can be deduced from gravitational wave measurements.

To describe the computational framework, we have focused on simple choices 
such as our ansatz neural ODE model Eq.~\eqref{eq:UDEModel} and the quadrupole formula for computing the harmonic modes.
These modeling choices are distinct from the overall computational framework and can be easily swapped out for different choices. For example, future high-accuracy studies should seek better orbit-to-waveform maps as the 
quadrupole formula is likely a large source of systematic error.

Our framework for learning the dynamics of binary black holes is quite general, and we expect that it can be applied to a variety of cases we have not considered including unequal masses, aligned-spins systems, and precessing systems.
Our method should be especially useful for discovering equations of motion for systems
where traditional approaches are less well-developed including
eccentric binaries, the highly relativistic late-inspiral and plunge dynamical regimes, and beyond-GR 
theories. Given the close agreement with NR trajectories, other possible applications could include setting NR initial data whereby the neural ODEs could be used to predict an NR trajectory before the simulation is performed.

One of the most important applications of our technique may involve calibrating existing orbital dynamics models (and high-accuracy gravitational-wave models) by using a base model different from Eq.~\eqref{eq:UDEModel}. Given that all modern inspiral-merger-ringdown waveform models require calibration of unknown parameters to numerical relativity simulations, the waveform inversion technique described here could benefit these efforts. For example, if implemented within the effective-one-body (EOB) approach, a suitably modified version of our methodology could be used to calibrate for missing terms in the EOB Hamiltonian. Precessing NR surrogate models also require a dynamical model, which is found through a direct fitting for the right-hand-side of the relevant ODEs. This computationally costly step might be simplified and accelerated with our techniques.

Another potential use of our methodology is training dynamical models entirely from gravitational-wave datasets instead of solving or analyzing Einstein's equation of general relativity as is traditionally done. In Fig.~\ref{fig:vsSNR} we explore how well the algorithm performs as the signal's SNR is systematically varied. We find that, at least for the examples considered here, the method continues to work down to an SNR of about 32. Consequently, our method is most applicable for future GW detections, including those made with LISA~\cite{2017arXiv170200786A}, Cosmic Explorer~\cite{Evans:2016mbw}, Einstein Telescope~\cite{Hild:2010id}, or perhaps the upcoming LIGO-Virgo-Kagra science run. Our method will need to be modified to achieve SNR levels of approximately 10 (which would cover most GW detections to date~\cite{abbott2020gwtc}) without compromising accuracy. Some approaches could include: (i) comparing waveforms with the Wasserstein metric which is known to be more robust to phase trappings, which shows up in higher noise, (ii) detecting an ensemble of noisy signals, and training on this observation set, (iii) using Bayesian networks where the network parameters are probability functions, or (iv) to apply our technique to filtered waveforms using a model-agnostic approach such as wavelet methods or denoising methods. We leave such extensions for future work.

\bigskip
\begin{acknowledgments}
We thank Alvin Chua, Sebastian Khan, Gaurav Khanna, Tom O'Leary-Roseberry, and Harald Pfeiffer for helpful discussions.
S.E.F.\ is supported by NSF grants No. PHY-1806665, PHY-2110496, and No. DMS-1912716.
This manuscript was written while the authors were in residence at the Institute for Computational and Experimental Research in Mathematics (ICERM) in Providence, RI, during the Advances in Computational Relativity program, supported by the National Science Foundation under Grant No. DMS-1439786.

This work was performed under the auspices of the U.S. Department of Energy by Lawrence Livermore National Laboratory under Contract DE-AC52-07NA27344, LLNL-JRNL-819108.
\end{acknowledgments}

\appendix
\section{Methods} \label{sec:methods}

In this section, we elaborate on the technical elements of our work which are necessary for replication of the results.
We open with a brief overview of the adjoint sensitivity method \cite{boltyanskiy1962mathematical} which we used to compute derivatives of $\mathcal{J}$ (see Eqs.~\cref{eq:Loss,eq:NewJ}) with respect to $\bm{\xi}$ and, in turn, facilitate solving problem~\cref{eq:InverseProblem}.
The section then closes with an overview of the quadrupole formula we have used to model the gravitational waveform~\cref{eq:gw_22_v2}.
Because both of these topics are well-known in specific (but mainly disjoint) scientific communities, we keep the exposition brief but include numerous references to the literature.

\subsection{Calculation of derivatives} \label{sub:implementation}

The ODE-constrained optimization problem~\Cref{eq:InverseProblem} delivers a calibrated dynamical system model
\begin{equation}
\label{eq:ForwardProblem}
	\dot{\mathbf{x}} = \mathbf{F}(\mathbf{x};{\bm{\xi}})
		,
	\qquad
	\mathbf{x}(0) = \mathbf{x}_0
	,
\end{equation}
where $\mathbf{F}(\mathbf{x};{\bm{\xi}}) = \mathbf{f}(\mathbf{x},\mathcal{F}(\mathbf{x};{\bm{\xi}}))$.
Because $\mathbf{x} = \mathbf{x}(\bm{\xi})$ depends implicitly on $\bm{\xi}$ through Eq.~\cref{eq:ForwardProblem}, the main technical difficulty with solving such optimization problems lies in computing total derivatives of the functional
\begin{equation}
\label{eq:AbstractFunctional}
	\mathcal{J}(\mathbf{x},\bm{\xi})
	=
	\frac{1}{T}
	\int_{0}^{T}
	J(\mathbf{x},{\bm{\xi}})
	\,
	\mathrm{d}t
	,
\end{equation}
with respect to $\bm{\xi}$.

Indeed, assume that we are working with the definition of $\mathcal{J}$ given in Eq.~\cref{eq:Loss}.
Here, any gradient-based optimization algorithm requires the calculation of the total derivative
\begin{equation}
	\mathrm{d}_{\bm{\xi}} \mathcal{J}
	=
	\frac{1}{T}
	\int_{0}^{T}
	\partial_{\mathbf{x}} J \mathrm{d}_{\bm{\xi}} \mathbf{x}
		+
	\partial_{\bm{\xi}} J
		\,
	\mathrm{d}t
	.
\end{equation}
One may easily note that, in the specific setting given to us through Eq.~\cref{eq:Loss}, we have $\mathcal{J} = \mathcal{J}(\mathbf{x})$ and, therefore, the partial derivative $\partial_{\bm{\xi}} J$ vanishes.
In more general situations, the term $\partial_{\bm{\xi}} J$ is routine to derive.
On the other hand, the calculation of $\mathrm{d}_{\bm{\xi}} \mathbf{x}$ is problematic due to the fact that $\mathbf{x}(\bm{\xi})$ is not available in closed form.
One approach to computing $\mathrm{d}_{\bm{\xi}} \mathcal{J}$ involves directly estimating $\mathrm{d}_{\bm{\xi}} \mathbf{x}$ via finite differences, however, the cost of this approach scales linearly with the size of $\bm{\xi}$.
This makes it prohibitive for most practical problems.
We choose to compute these gradients using in an alternative way often referred to as the adjoint sensitivity method \cite{boltyanskiy1962mathematical}.

The adjoint sensitivity method has been used extensively in engineering design \cite{jameson1988aerodynamic,giles2000introduction} and, more recently, in machine learning research \cite{chen2018neural,rackauckas2020universal,belbute2020combining}.
It involves the integration of two ODEs over the time interval $[0,T]$: the original governing ODE~\cref{eq:ForwardProblem} and an adjoint ODE (integrated backwards in time).

The derivation of this method usually begins with the Lagrangian
\begin{equation}
\begin{aligned}
	\mathcal{L}
		&=
	\frac{1}{T}
	\int_{0}^{T}
	J(\mathbf{x}(t),\bm{\xi}) - \bm{\lambda}(t)^\top\big(\dot{\mathbf{x}}(t) - \mathbf{F}(\mathbf{x}(t);{\bm{\xi}}))\big)
	\,
	\mathrm{d}t
	\\
	&\phantom{=}-
		\bm{\mu}^\top \big(\mathbf{x}(0) - \mathbf{x}_0\big)
	,
\end{aligned}
\end{equation}
which comes from writing the $\bm{\xi}$-minimization of~\cref{eq:AbstractFunctional}, constrained by solutions of~\cref{eq:ForwardProblem}, as a saddle-point problem \cite{ekeland1999convex}.
This functional is clearly designed such that $\mathrm{d}_{\bm{\xi}} \mathcal{J} = \mathrm{d}_{\bm{\xi}} \mathcal{L}$ for any $\mathbf{x}$ satisfying~\cref{eq:ForwardProblem}.
In addition, one observes that
\begin{align}
			\mathrm{d}_{\bm{\xi}} \mathcal{L}
	&=
	\partial_{\mathbf{x}} \mathcal{L}\, \mathrm{d}_{\bm{\xi}}\mathbf{x}
	+
	\partial_{\bm{\lambda}} \mathcal{L}\, \mathrm{d}_{\bm{\xi}}\bm{\lambda}
	+
	\partial_{\bm{\mu}} \mathcal{L}\, \mathrm{d}_{\bm{\xi}}\bm{\mu}
	+
	\partial_{\bm{\xi}} \mathcal{L}
	\\
	&=
	\frac{1}{T}
	\int_{0}^{T}
	\partial_{\bm{\xi}} J(\mathbf{x})
	-
	\bm{\lambda}^\top \partial_{\bm{\xi}} \mathbf{F}(\mathbf{x};{\bm{\xi}})
	\,
	\mathrm{d}t
	,
\end{align}
if $\partial_{\bm{\lambda}} \mathcal{L} = \partial_{\bm{\mu}} \mathcal{L} = \partial_{\mathbf{x}} \mathcal{L} = 0$.
A straight\-forward calculation shows that the first two of these equations are equivalent to the original dynamical system~\cref{eq:ForwardProblem}.
On the other hand, $\partial_{\mathbf{x}} \mathcal{L} = 0$ is equivalent to the \emph{adjoint equation}
\begin{align}
\label{eq:BackwardProblem}
	-\dot{\bm{\lambda}}
	=
	[\partial_\mathbf{x}\mathbf{F}(\mathbf{x};{\bm{\xi}})]^\top\bm{\lambda}
	+
	\partial_\mathbf{x} J(\mathbf{x},{\bm{\xi}})
	,
	\qquad
	\bm{\lambda}(T) = \bm{0}
	.
\end{align}

Evidently, the ODE system~\cref{eq:BackwardProblem} depends on the solution to~\cref{eq:ForwardProblem}, $\mathbf{x}(t)$.
Therefore, the algorithm for computing $\mathrm{d}_{\bm{\xi}} \mathcal{J}$ must follow a specific order:
\begin{enumerate}
	\item
	Integrate $\dot{\mathbf{x}} = \mathbf{F}(\mathbf{x};{\bm{\xi}})$ from $t = 0$ to $T$ with the initial condition $\mathbf{x}(0) = \mathbf{x}_0$.
	\item
	Integrate $-\dot{\bm{\lambda}} = [\partial_\mathbf{x}\mathbf{F}(\mathbf{x};{\bm{\xi}})]^\top\bm{\lambda} + \partial_\mathbf{x} J(\mathbf{x},{\bm{\xi}})$ from $t = T$ to $0$ with the initial condition $\bm{\lambda}(T) = \bm{0}$.
	\item
	Compute $\mathrm{d}_{\bm{\xi}} \mathcal{J} = \frac{1}{T}\int_{0}^{T}\partial_{\bm{\xi}} J(\mathbf{x})-\bm{\lambda}^\top \partial_{\bm{\xi}} \mathbf{F}(\mathbf{x};{\bm{\xi}})\,\mathrm{d}t$.
\end{enumerate}
Extension of this algorithm to the scenario where the initial condition $\mathbf{x}_0$ is also optimized for (as considered in, e.g., \Cref{ssub:low_eccentricity_orbit}) is straightforward.
For thorough accounts of the numerical implementation of the adjoint sensitivity method for UDEs, we refer the interested reader to \cite{rackauckas2020universal,Gholami2020ANODE}.

\subsection{Gravitational waves from an orbit}\label{sub:gravitational_waves_from_an_orbit}

In the context of general relativity,
gravitational waves are associated with the outgoing, radiative parts of the 
spacetime metric and are solutions to Einstein field 
equation. The motion of massive objects produce gravitational waves, and 
our model outlined in Sec.~\ref{sec:physically_motivated_models} provides the equations of motion for
object 1 of mass $m_1$ and position $\mathbf{r}_1(t)$, and 
object 2 of mass $m_2$ and position $\mathbf{r}_2(t)$.

The quadrupole formula provides one simple method of obtaining the gravitational radiation from these orbital trajectories. 
In this framework, we assume 
both black holes to be ``point sources'' (i.e. Dirac delta functions). The Newtonian mass density of two objects orbiting in the x-y plane is 
\begin{align}
\rho(t,x,y,z) & = m_1 \delta(z) \delta(x - x_1(t) ) \delta(y - y_1(t) ) \nonumber \\
  & + m_2 \delta(z) \delta(x - x_2(t) ) \delta(y - y_2(t) ) \,,
\end{align}
and note that
in the special case $m_2 \gg m_1$ we have $\mathbf{r}_2(t) = (0,0,0)$ and
$\mathbf{r}_1(t) =  \mathbf{r}(t)$. Given the density, the quadrupole formula tells us
that the dominant quadrupole mode of the gravitational radiation field tensor
in the transverse-traceless gauge is 
\begin{align} \label{eq:gw}
r H^{ab} = 2 \frac{\partial^2}{\partial t^2} \mathcal{I}^{ab} \,,
\end{align}
where the trace-free mass quadrupole tensor is
\begin{align} \label{eq:gw_mathcalI}
{\cal I}^{ab} 
= I^{ab} - \frac{1}{3} \delta^{ab}\delta_{cd}I^{cd} \,,
\end{align}
$\delta^{ab}$ is the Kronecker delta. 
In Cartesian coordinates, the indicies take on values of ``x", ``y", and ``z".
For example, we have $H^{xx}$, $H^{yy}$ , $H^{xy}$, etc.
The components of the mass quadrupole tensor, $I^{ab}$, that are relevant to Eq.~\eqref{eq:gw} (non-zero temporal derivatives) are
\begin{subequations}
\label{eq:I}
\begin{align}
	I^{xx} & = \int d^3x \rho x^2 = m_1 x_1(t)^2 + m_2 x_2(t)^2  \,, \\	I^{yy} & = \int d^3x \rho y^2 = m_1 y_1(t)^2 + m_2 y_2(t)^2 \,, \\
	I^{xy} & = \int d^3x \rho xy = m_1 x_1(t)y_1(t) + m_2 x_2(t)y_2(t) \,,
\end{align}
\end{subequations}
and by symmetry $I^{xy} = I^{yx}$. 

This framework computes the gravitational wave as perturbations, $H^{ab}$, of the background spacetime metric. However, 
numerical simulations instead provide the plus, $h_+$ and cross, $h_{\times}$, gravitational-wave polarizations
defined on a ``sphere at infinity", which are obtained after
contracting $H^{ab}$ with the polarization tensors~\cite{bishop2016extraction}. 
On this sphere, it is common to define a 
complexified gravitational wave
\begin{align}
h(t, \theta, \phi)  & =  h_+(t, \theta, \phi) - \mathrm{i}  h_{\times}(t, \theta, \phi) \\
& = \sum^{\infty}_{\ell=2} \sum_{m=-l}^{l}
        h^{\ell m}(t) ~_{-2}Y_{\ell m}(\theta, \phi) \,,
\end{align}
and subsequently expand $h$ into a complete
basis of spin$\,=\!\!-2$ weighted spherical harmonics labeled by $(\ell, m)$
harmonic indices, $_{-2}Y_{\ell m}$.
Here $\theta$ and $\phi$ are the polar and azimuthal angles.
For example, the SXS $(2,2)$ mode gravitational waveform data, $h^{22}$, was used extensively in this paper.
Given the orbital trajectories, the $(2,2)$ mode,
\begin{align}
\label{eq:h22_appendix}
h^{22}(t) &=\dfrac{1}{r}\sqrt{\dfrac{4\pi}{5}}\left(\ddot{\mathcal{I}}_{xx}
- 2 i\ddot{\mathcal{I}}_{xy}-\ddot{\mathcal{I}}_{yy}\right)\,,
\end{align}
can be computed directly from the trace-free mass quadrupole tensor~\cite[Eqs.~54--56]{bishop2016extraction}.

To summarize, from the orbital motion we compute the three components of the
mass quadrupole tensor Eq.~\eqref{eq:I}, compute the trace-free mass quadrupole tensor, compute the time derivatives using finite differences, and then finally assemble the (2,2)-multipolar component of the outgoing gravitational waves, $h^{22}$, from Eq.~\eqref{eq:h22_appendix}.

While a full discussion is outside the
scope of this appendix, we point out that the quadrupole formula is the simplest possible one
and, consequently, ignores a lot of the relevant physics. Future work could consider incorporating more physics into the gravitational waveform model, including
relativistic definitions of the density, higher-order post-Minkowskian corrections, subdominant
harmonic modes, or
near-field-to-far-field transport maps.
Nevertheless, some of missing physics might already be accounted for through the orbital 
model, where the deep networks may try to account for  missing physics in the waveform model by modifying
the orbital dynamics model.

\phantomsection\bibliography{main}

\end{document}